\documentclass[preprint]{aastex63}

\newcommand{\goes}{{\em GOES}}
\newcommand{\lya}{Ly$\alpha$}

\submitjournal{ApJ}

\shorttitle{The Lyman-alpha Emission in Solar Flares. I.}
\shortauthors{Jing et al.}

\begin{document}

\title{The Lyman-alpha Emission in Solar Flares. I. a Statistical Study \\on Its Relationship with the 1--8 \AA\ Soft X-ray Emission}
\author{Zhichen Jing}
\affiliation{School of Astronomy and Space Science, Nanjing University, Nanjing 210023, People's Republic of China}
\author{Wuqi Pan}
\affiliation{School of Astronomy and Space Science, Nanjing University, Nanjing 210023, People's Republic of China}
\author{Yukun Yang}
\affiliation{School of Astronomy and Space Science, Nanjing University, Nanjing 210023, People's Republic of China}
\author{Dechao Song}
\affiliation{Key Laboratory of Dark Matter and Space Astronomy, Purple Mountain Observatory, Chinese Academy of Sciences, Nanjing 210033, People's Republic of China}
\affiliation{School of Astronomy and Space Science, University of Science and Technology of China, Hefei 230026, People's Republic of China}
\author{Jun Tian}
\affiliation{Key Laboratory of Dark Matter and Space Astronomy, Purple Mountain Observatory, Chinese Academy of Sciences, Nanjing 210033, People's Republic of China}
\affiliation{School of Astronomy and Space Science, University of Science and Technology of China, Hefei 230026, People's Republic of China}
\author{Y. Li}
\affiliation{Key Laboratory of Dark Matter and Space Astronomy, Purple Mountain Observatory, Chinese Academy of Sciences, Nanjing 210033, People's Republic of China}
\affiliation{School of Astronomy and Space Science, University of Science and Technology of China, Hefei 230026, People's Republic of China}
\author{X. Cheng}
\affiliation{School of Astronomy and Space Science, Nanjing University, Nanjing 210023, People's Republic of China}
\author{Jie Hong}
\affiliation{School of Astronomy and Space Science, Nanjing University, Nanjing 210023, People's Republic of China}
\author{M. D. Ding}
\affiliation{School of Astronomy and Space Science, Nanjing University, Nanjing 210023, People's Republic of China}

\correspondingauthor{Ying Li}
\email{yingli@pmo.ac.cn}

 \begin{abstract}
We statistically study the relationship between the Lyman-alpha (\lya) and 1--8 \AA\ soft X-ray (SXR) emissions from 658 M- and X-class solar flares observed by the {\em Geostationary Operational Environmental Satellite} during 2006--2016. Based on the peak times of the two waveband emissions, we divide the flares into three types. Type I (III) has an earlier (a later) peak time in the \lya\ emission than that in the SXR emission, while type II has nearly a same peak time (within the time resolution of 10 s) between the \lya\ and SXR emissions. In these 658 flares, we find that there are 505 (76.8\%) type I flares, 10 (1.5\%) type II flares, and 143 (21.7\%) type III flares, and that the three types appear to have no dependence on the flare duration, flare location, or solar cycle. Besides the main peak, the \lya\ emission of the three type flares also shows sub-peaks which can appear in the impulsive or gradual phase of the flare. It is found that the main-peak (for type I) and sub-peak (for type III) emissions of \lya\ that appear in the impulsive phase follow the Neupert effect in general. This indicates that such \lya\ emissions are related to the nonthermal electron beam heating. While the main-peak (for type III) and sub-peak (for type I) emissions of \lya\ that appear in the gradual phase are supposed to be primarily contributed by the thermal plasma that cools down.
\end{abstract}



\section{Introduction}
\label{sec:intro}

Solar flares are sudden brightenings and energetic events in the solar atmosphere \citep[e.g.,][]{flet11,shib11}. In the standard flare model (also called CSHKP model; \citealt{carm64,stur66,hira74,kopp76}), the magnetic energy is released in the corona through magnetic reconnection. The energy is then transported downward to the chromosphere via nonthermal electron beams and/or thermal conduction. Consequently, the chromospheric plasma is heated and goes up into the corona due to a thermal pressure (i.e., chromospheric evaporation). The evaporated hot plasma fills the flare loops that can be well seen in the soft X-ray (SXR) waveband. Based on the observed SXR light curve, the evolution of a flare can be classified into three phases, pre-flare phase, impulsive (or rise) phase, and gradual (or decay) phase (e.g., \citealt{flet11,huds11}). The SXR emission has begun to increase in the pre-flare phase; then it has a rapid rise, reaching its maximum in the impulsive phase because of a significant heating; in the gradual phase, it decays gradually due to the plasma cooling. The heating and cooling processes can also be manifested in some other wavebands. For example, the nonthermal electron beam heating produces a hard X-ray (HXR) emission that resembles the time derivative of the SXR emission (i.e., Neupert effect; \citealt{neup68}). While the hot plasma that is heated earlier will cools down gradually and be observed in the SXR and extreme-ultraviolet (EUV) wavebands successively.

In the solar UV spectrum, there is a strongest line of hydrogen Lyman-alpha (\ion{H}{1} \lya) at 1216 \AA\ \citep{curd01}. This line is formed in the mid-to-upper chromosphere and low transition region \citep{vern81}, which is optically thick and suffers an opacity effect (e.g., \citealt{vial82,wood95}). It has been reported that the \lya\ line shows a significant enhancement during solar flares \citep{wood04,rubi09,mill12,mill14,kret15,mill20}. However, up to now, we still have a poor understanding on this line and its physical origin in solar flares owing to relatively rare observations as well as a complex formation process of this line.

In the past decades, there have been a few studies on \lya\ in solar flares by using spectral/imaging/photometric observations. \cite{canf80} presented the \lya\ line profiles and their temporal variations for two flares observed by a slit spectrograph on {\em Skylab}. With the L.P.S.P. experiment on {\em OSO-8}, \cite{lema84} preformed a simultaneous observation of a flare in six chromospheric lines including \lya\ and found an indication of a downward energy propagation via the temporal behaviour of the different lines. The \lya\ emission was found to increase by about 0.6--30\% during flares \citep{brek96,wood04,kret13,mill20} with its radiated energy rate estimated to be $\sim$10$^{25-27}$ erg s$^{-1}$ \citep{john11,mill12,mill14}. In particular, \cite{mill20} carried out a statistical study based on 477 M- and X-class flares observed by the {\em Geostationary Operational Environmental Satellite} (\goes) and showed that 95\% of these major flares have an enhancement of 10\% or less in \lya, with a maximum of about 30\% in all of the flares. The authors also reported that there is a center-to-limb variation in the \lya\ emission due to an opacity effect. Besides \lya\ enhancements, quasi-periodic behaviours at one or three minutes were detected in the \lya\ emission during flares \citep{mill17,lido20}. In addition, there were some studies focusing on the relationship of the \lya\ emission with soft X-ray (SXR) and hard X-ray (HXR) emissions. \cite{nusi06} reported that the \lya\ maximum is reached well before the SXR maximum and the variation of the \lya\ emission is synchronous with that of the HXR emission above 50 keV during the impulsive phase of a flare. These ``Neupert-effect" features were also observed and confirmed by some other authors via studying different flare events \citep{rubi09,mill16,mill17,domi18,cham18}. In particular, with the \lya\ imaging observations from the {\em Transition Region and Coronal Explorer} ({\em TRACE}), \cite{rubi09} demonstrated that most of the \lya\ emission is co-spatial with the HXR sources which originate from the flare footpoints. These observational results indicate that the \lya\ emission in flares has a nonthermal origin.

Some modelling works on flaring \lya\ have also been emerging in quite recent years. Using the radiative hydrodynamics code RADYN, \cite{brow18} simulated the response of the \lya\ line with hard and soft electron beam heatings and compared the synthetic \lya\ line profiles with the observed ones from the Extreme-ultraviolet Variability Experiment (EVE) onboard the {\em Solar Dynamics Observatory} ({\em SDO}). Via RADYN, \cite{hong19} also calculated the \lya\ line profiles in nonthermal and thermal heating models in which the \lya\ line exhibits different features in line asymmetry and light curves. Moreover, \cite{yang20} modelled the \lya\ and H$\alpha$ lines as well as the continua at 3600 \AA\ and 4250 \AA\ in white-light flares and found that the \lya\ line has different responses to the nonthermal beam heating with the H$\alpha$ line and the continua. Besides RADYN, the HYDRO2GEN code was used to calculate the Lyman lines and continuum with a beam injection model \citep{drue19}. These simulations show that the \lya\ line has a notable response to the nonthermal and also thermal heatings.

In order to study the physical origin of the \lya\ emission in solar flares, here we statistically study the relationship between the \lya\ and SXR emissions from 658 M- and X-class flares observed by \goes\ during 2006--2016. We find that in most of the flares, the \lya\ emission peaks earlier than the SXR emission and holds the Neupert effect in general, indicative of a nonthermal origin, just as previous studies demonstrated. However, we also find that there are about one fifth flares with their \lya\ peak later than the SXR peak, which may primarily have a thermal origin. To the best of our knowledge, the latter result has never been reported in observations before, as least from a statistical aspect. The statistical study in the present work can improve our understanding on the physical origin of the \lya\ emission in solar flares. In the following, we describe the instruments and data in Section \ref{sec:data} and our flare dataset in Section \ref{sec:flare}. The statistical results are shown in Section \ref{sec:result}, followed by a summary and interpretation in Section \ref{sec:summary}. In the last Section \ref{sec:dis}, we give the conclusion and discussions.


\section{Instruments and Data}
\label{sec:data}

The \goes\ series spacecraft (\goes-1--17) have been providing the solar X-ray irradiance continuously since 1975. The X-ray Sensor (XRS; \citealt{hans96}) on \goes\ observes the full-disk Sun in two soft X-ray channels, i.e., 0.5--4 \AA\ (short channel) from XRS-A and 1--8 \AA\ (long channel) from XRS-B, the latter of which is widely used to define the flare magnitude from A- to X-class. Since \goes-13 (launched in 2006), there also includes an Extreme Ultraviolet Sensor (EUVS; \citealt{vier07}) that observes the whole Sun in the EUV wavebands. EUVS contains five channels, called A, B, C, D, E, from 50--1270 \AA\ with the E channel targeting the \lya\ emission. The E channel has a width of $\sim$90 \AA\ covering a wavelength range of 1180--1270 \AA, whose emission primarily comes from the \lya\ line. 

In this work, we mainly use the \lya\ and 1--8 \AA\ SXR data. These data have been converted into irradiances in units of W m$^{-2}$, which have cadences of $\sim$10 s and 2 s, respectively. The \lya\ data that have been publicly released are for the period of 2006--2016 (the \goes-13--15 era), namely covering an entire solar cycle. Considering that some \lya\ data for a certain time were simultaneously obtained by two or even three spacecraft, here we choose the data on the following basis: from \goes-15 first, if no then from \goes-14, and finally from \goes-13. In order to obtain the flare-induced emission in both of the \lya\ and SXR wavebands, we have made a background subtraction by averaging the pre-flare flux from one hour before the flare onset. Note that a correction (divided by 0.7) for the ``true" SXR flux is not applied in this study, which is followed by the work in \cite{mill20}. In addition to the \lya\ and SXR data from \goes, we use the imaging data from the Atmospheric Imaging Assembly (AIA; \citealt{leme12}) on board {\em SDO} and the sunspot data\footnote{http://sidc.oma.be/silso/datafiles} from the World Data Center SILSO at the Royal Observatory of Belgium in Brussels.


\section{Flare Dataset}
\label{sec:flare}

\goes\ recorded more than 700 X- and M-class flares during the years of 2006--2016. Here we have 658 flare examples, 43 (6.5\%) X-class flares and 615 (93.5\%) M-class flares, among them in our dataset. We have excluded the flare events with bad or poor data as well as the ones that are affected by geocoronal absorption in \lya. Below list some information on our flare dataset.

\begin{figure}[htb]
\centering
\includegraphics[scale = 0.39]{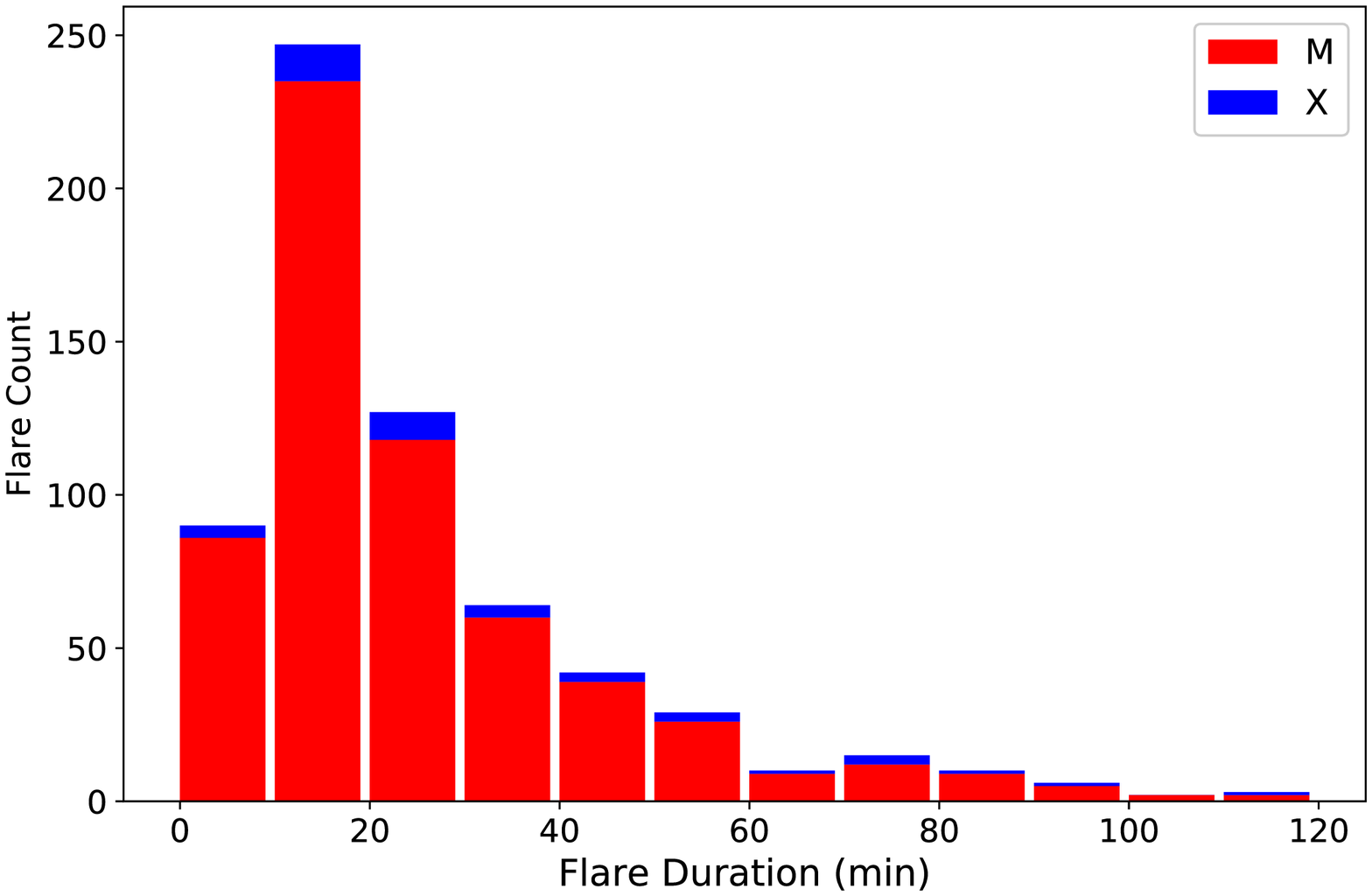}
\includegraphics[scale = 0.39]{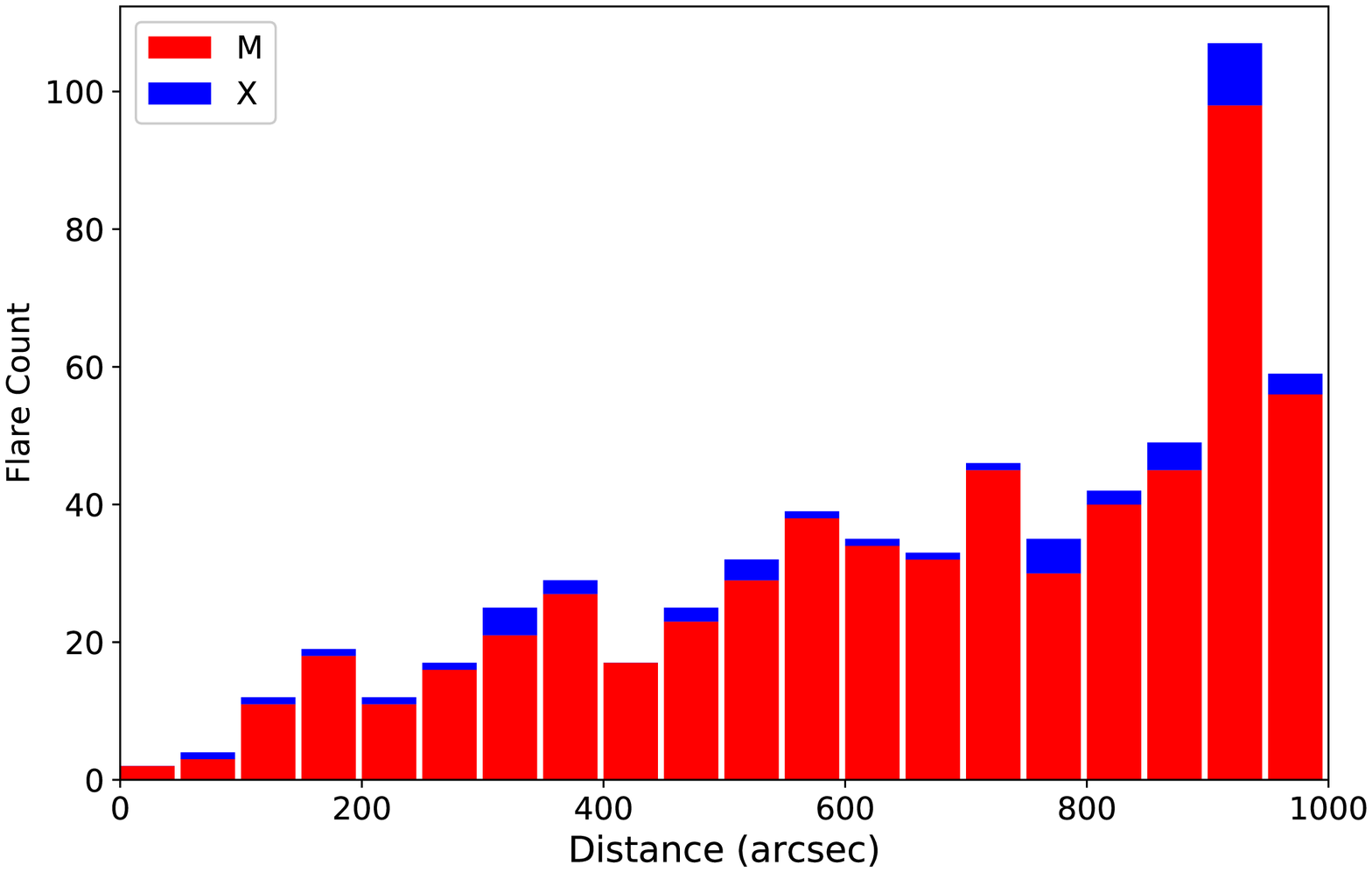}\\
\includegraphics[scale = 0.395]{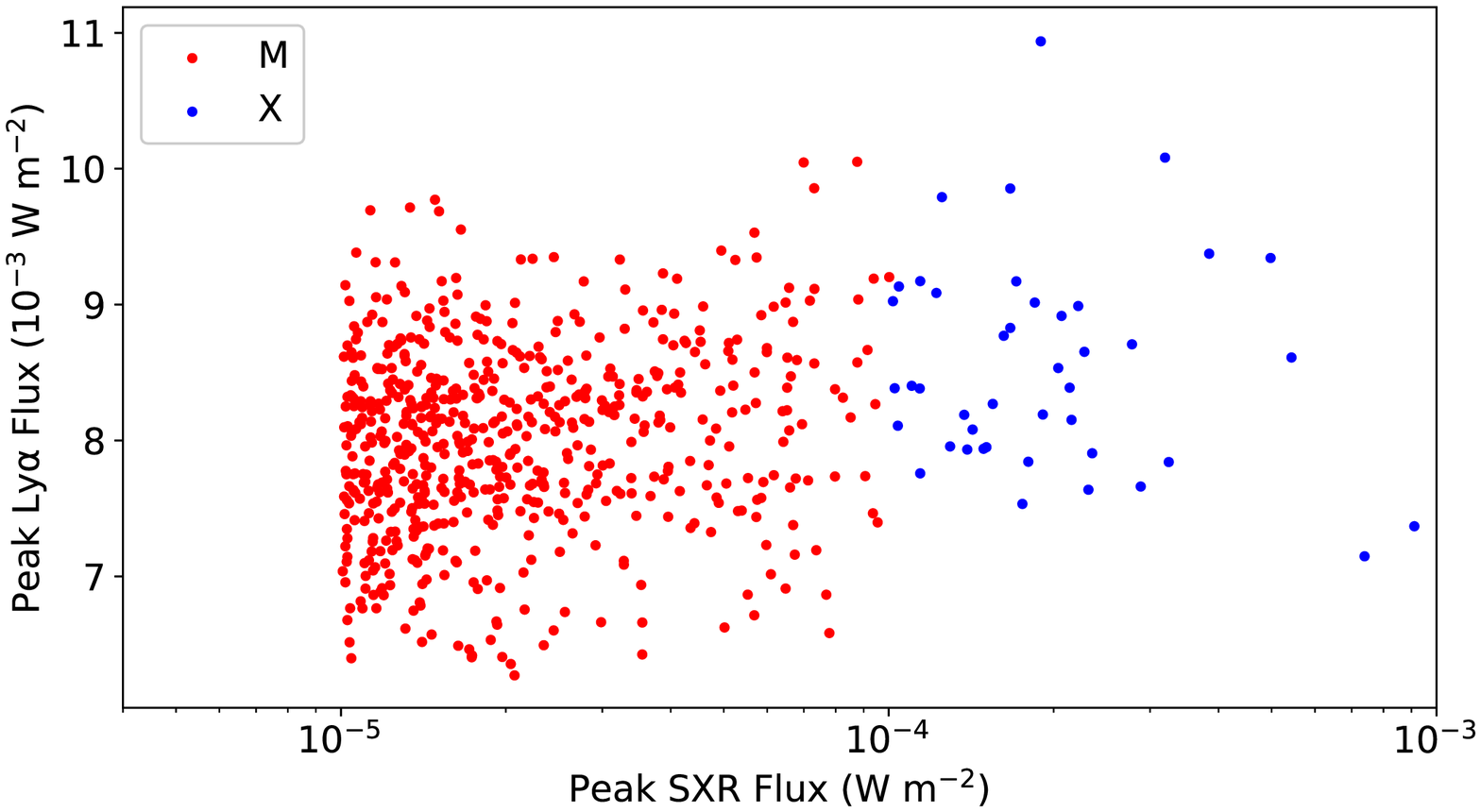}
\includegraphics[scale = 0.395]{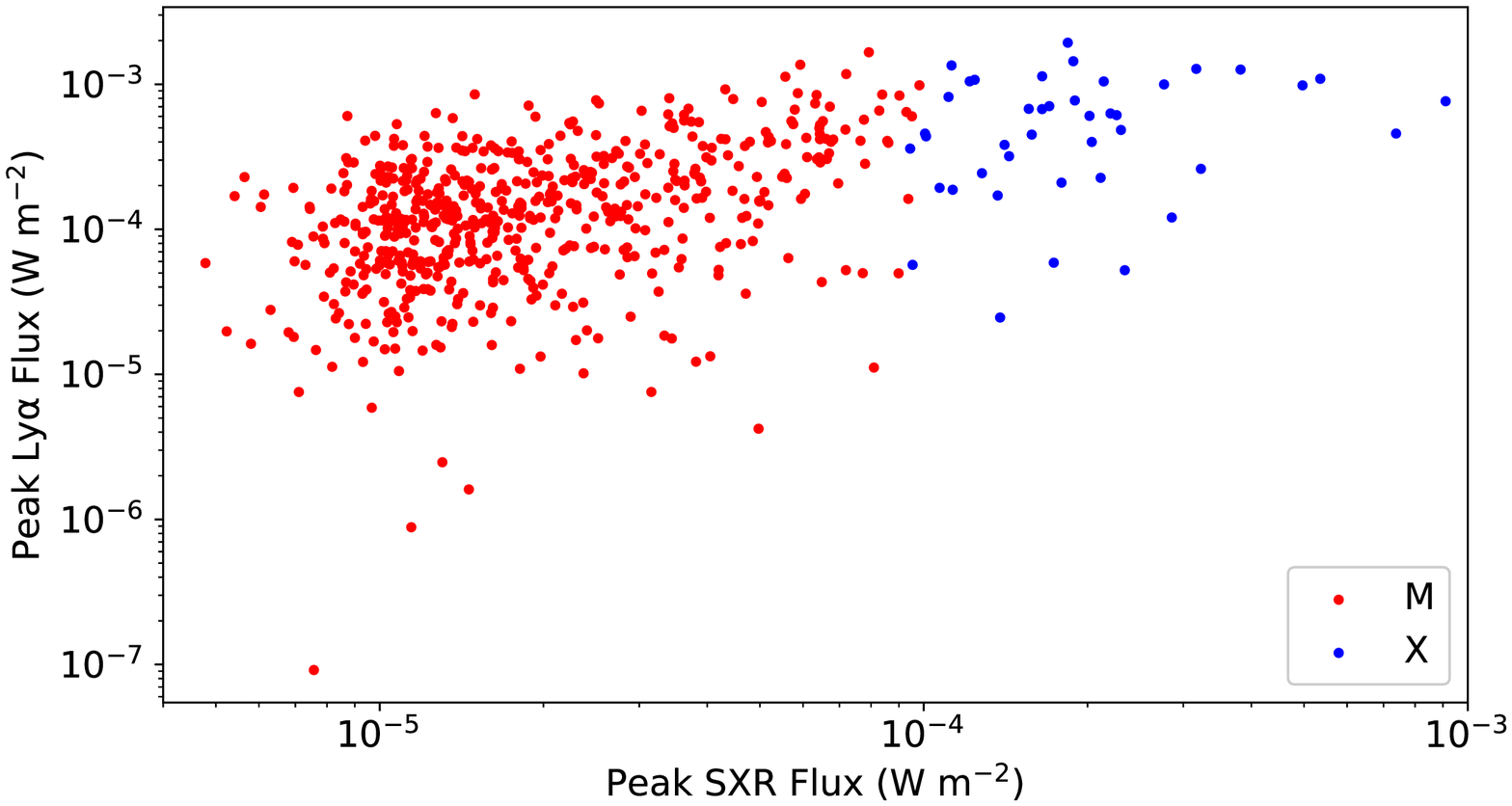}
\caption{Top panels show the histograms of flare duration (left) and flare distance from disk center (right). Note that the duration histogram has been cut to 120 min for a better display. Also note that the flare distance suffers a foreshortening effect and the increasing trend from disk center to the limb does not reflect a real distribution of flares per surface area. Bottom panels are the scatter plots of peak \lya\ flux versus peak SXR flux, with the left one without a background subtraction and the right one with the background subtracted. In each of the panels, the red and blue colors mark the M- and X-class flares, respectively.}
\label{fig:tau}
\end{figure}

{\em Duration of the flares.} The distribution histogram of the flare duration is given in the top left panel of Figure \ref{fig:tau}, which is calculated from the start and end times recorded in the \goes\ flare list. Note that there are 13 ($\sim$2\%) flares that last for $>$120 minutes and are randomly distributed in the time range of 120--720 minutes and we do not show them in this histogram just for a better display. It is seen that the flare dataset contains both short- and long-duration events that last for a few minutes to hours, though most of the flare events end within 30 minutes. Note that \goes\ only records part of the decay phase of the flare, which actually underestimates the flare duration. It is also seen that the flare duration has no dependence on the flare magnitude, namely both short- and long-duration flares can be M- or X-class flares.

{\em Location of the flares.} We plot the distribution histogram of the flare location in the top right panel of Figure \ref{fig:tau}. The location is represented by the distance of the flare from disk center that is calculated from the X and Y coordinates\footnote{The X and Y coordinates are obtained from this webpage: https://helioportal.nas.nasa.gov/webapp.html}. Note that there are 19 ($\sim$3\%) M-class flares that lack the location information and thus are not included here. One can see that the flare examples in our dataset are located through the disk center (i.e., the distance is 0\arcsec) to the limb ($\sim$960\arcsec). Here it should be clarified that the increasing trend of the distribution is mainly caused by a foreshortening effect when the solar hemisphere is projected to the X--Y plane. This trend does not reflect the real distribution of the flare on the solar surface. In fact, according to the flare distribution as a function of heliocentric angle shown in \cite{mill20}, there exists a nearly uniform distribution of flares from disk center to the limb. The approximately uniform distribution can also be seen in the top right panel of Figure \ref{fig:dist} to some extent. Within the circle of a radius of 500\arcsec\ (or the heliocentric angle being $<$30 degrees more or less), where the foreshortening effect is somewhat trivial, we can see that the flare is in general distributed uniformly on the plane. From the histogram of flare location, one can also see that the location distribution has no dependence on the flare magnitude either, i.e., both X- and M-class flares can take place on the disk or at the limb.

{\em Peak fluxes of the flares.} In the bottom panels of Figure \ref{fig:tau}, we show the scatter plots of the peak \lya\ flux versus peak SXR flux for the flare examples, with the left one without subtracting the background and the right one with the background subtracted. It can be seen that there seems to be a correlation between the \lya\ and SXR peak fluxes whether the background is subtracted or not. We calculate the correlation coefficients for the two cases, which are 0.63 and 0.45, respectively. It should be mentioned that the pattern with subtracting the background is actually similar to the one as reported in \cite{mill20}, though we use a different method to find the background level. In the remaining paper, we just adopt the peak fluxes of both \lya\ and SXR that have been subtracted the background emission.


\section{Results}
\label{sec:result}

\begin{figure}[htb]
\centering
\includegraphics[scale = 0.8]{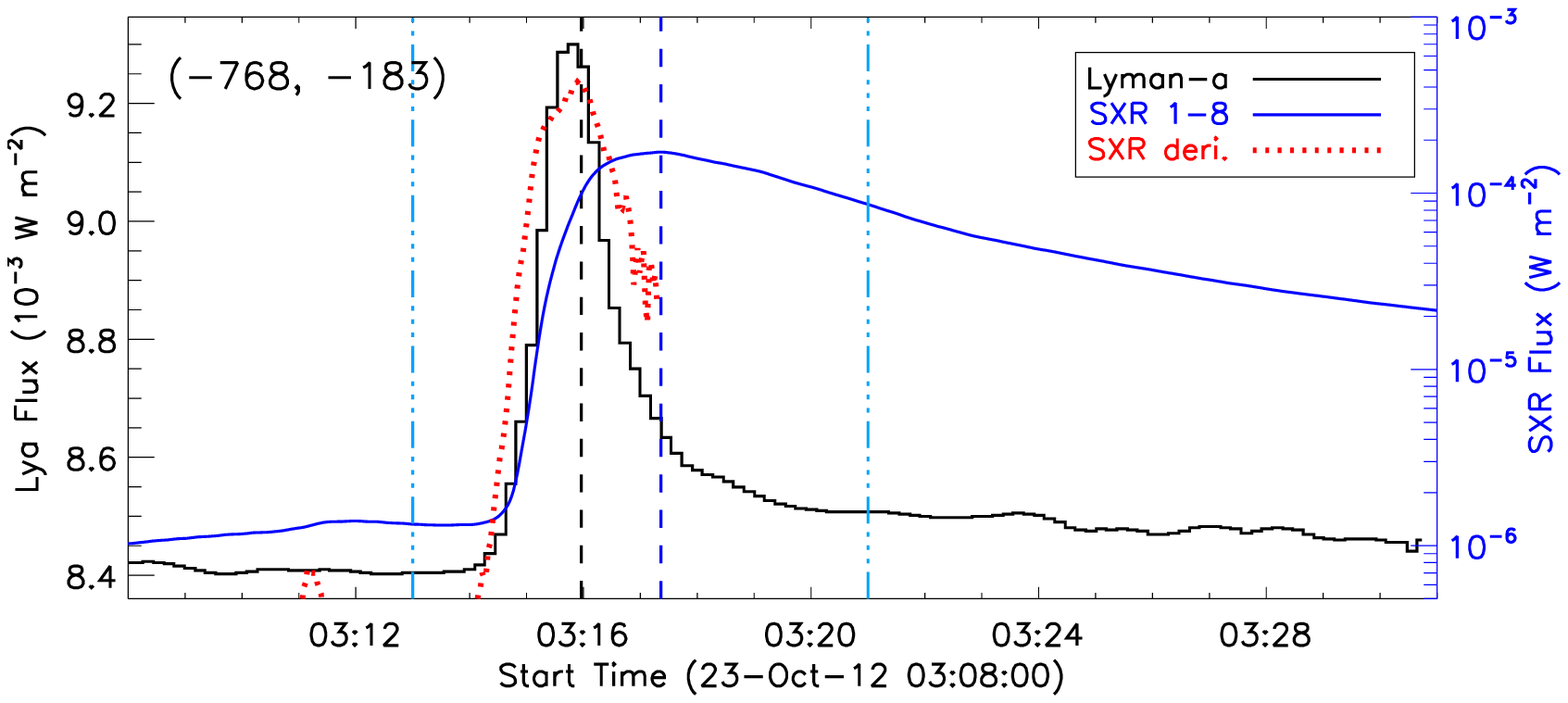}
\includegraphics[scale = 0.8]{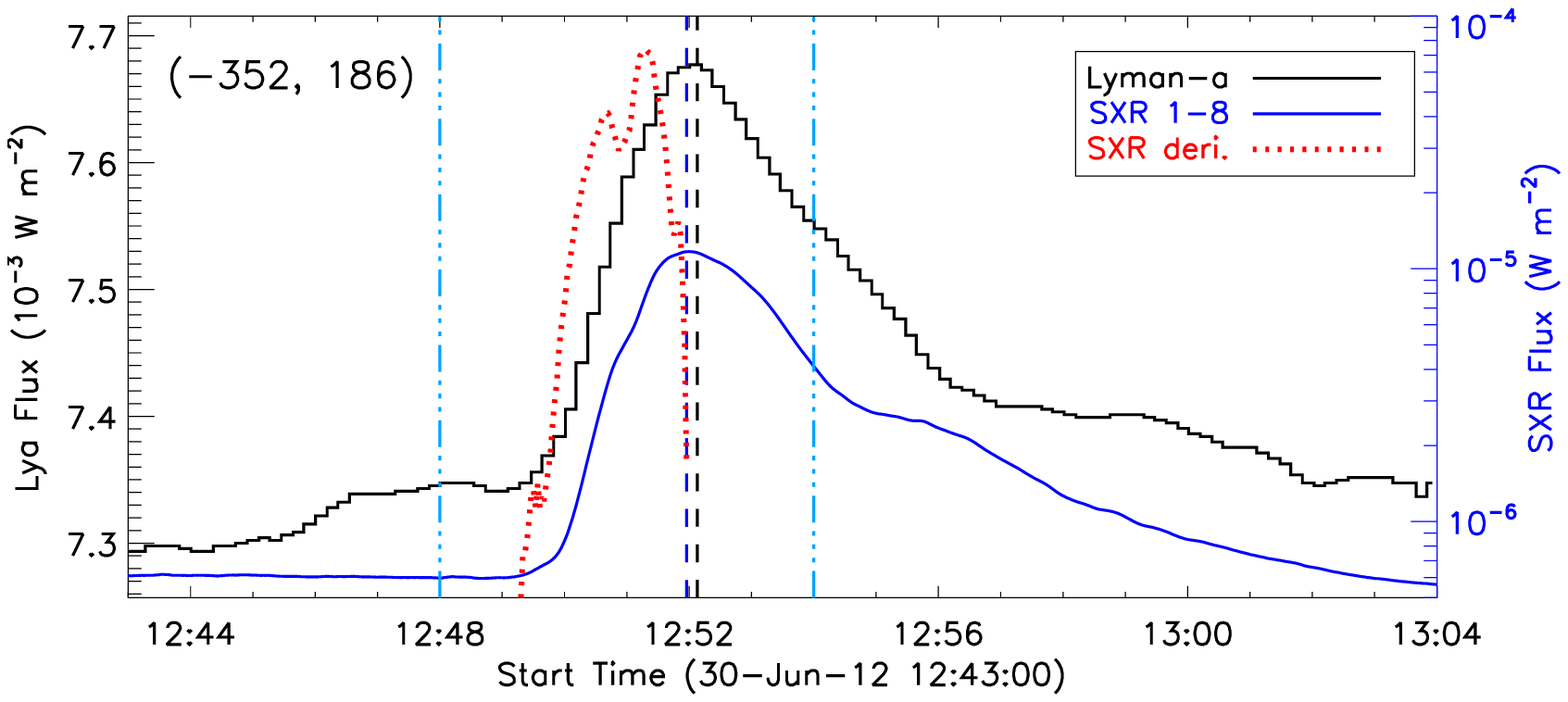}
\includegraphics[scale = 0.8]{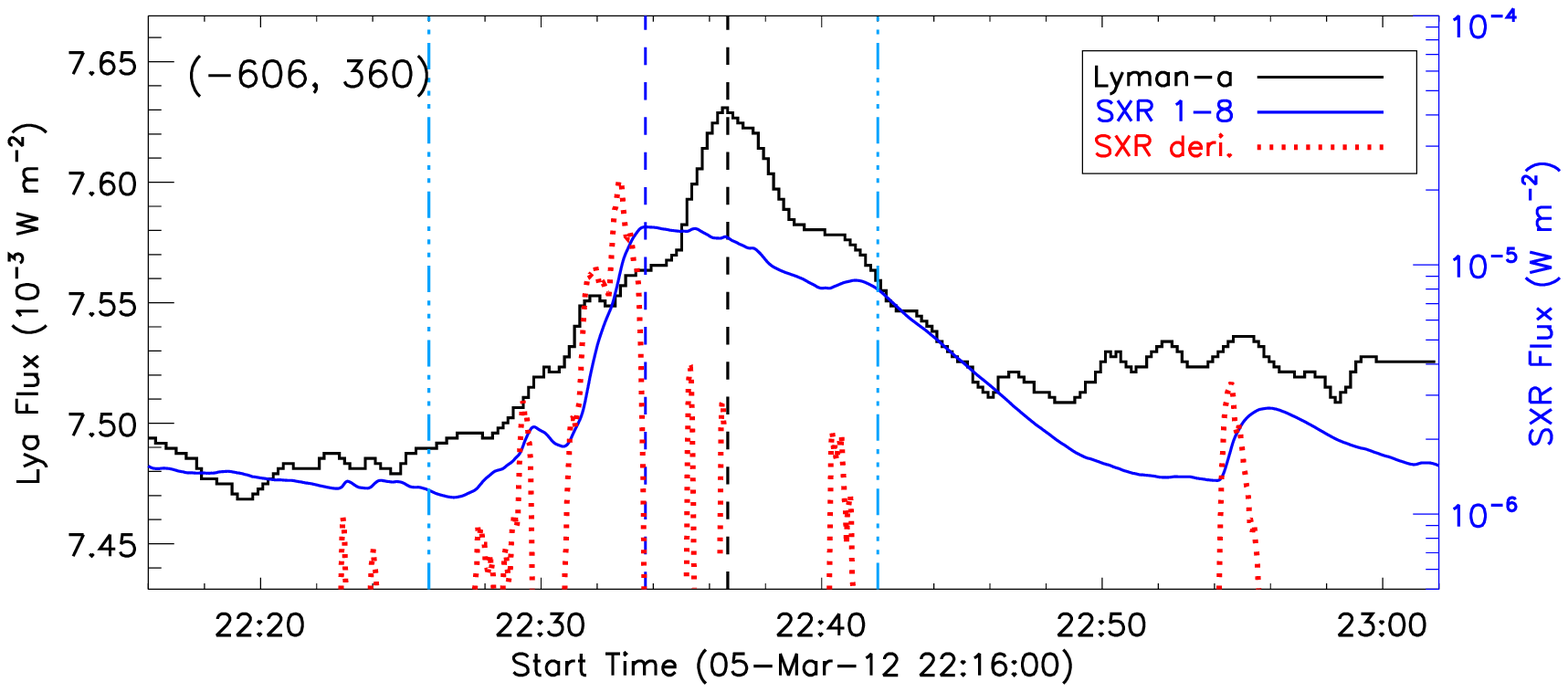}
\caption{Example light curves for the type I (top), type II (middle), and type III (bottom) flares. In each of the panels, the black curve denotes the \lya\ flux that corresponds to the left coordinate. The blue curve marks the 1--8 \AA\ SXR flux that corresponds to the right coordinate. The red dotted line represents the time derivative of the SXR flux in an arbitrary scale. The vertical black and blue dashed lines indicate the peak times of the \lya\ and SXR emissions, respectively. The two vertical light blue lines denote the start and end times of the flare recorded in the \goes\ flare list. In the top left corner, there gives the location of the flare in X and Y coordinates.}
\label{fig:types-lc}
\end{figure}

After viewing the \lya\ and SXR emission curves of all the flares in the dataset, we see some features as follows. (1) There is usually a rapid rise followed by a relatively slow decay in both of the \lya\ and SXR emission curves. (2) The \lya\ emission can reach its maximum (main peak) earlier or later than the SXR emission. (3) There appear two or even more evident peaks (main peak plus sub-peaks) in numerous flares particularly in the \lya\ emission curve. These features can be seen from the light curves as plotted in Figures \ref{fig:types-lc}, \ref{fig:typeI-lc1to4}, and \ref{fig:typeIII-lc1to4}.

\subsection{Three Types of Flares based on the Main Peak of the \lya\ emission}

\subsubsection{Example Light Curves}

According to the time sequence of the main peaks of \lya\ and SXR emissions, we divide the flares into three types: type I (III) has an earlier (a later) peak time in the \lya\ emission than that in the SXR emission, while type II has nearly a same peak time (within the time resolution, or $\pm$10 s) between the \lya\ and SXR emissions. In other words, the \lya\ emission in type I/III flares peaks in the impulsive/gradual phase of the flare, while the one in type II flares peaks around the same time as the SXR flux. These \lya\ emissions in different types are expected to be related to different physical processes according to the standard flare model. Figure \ref{fig:types-lc} gives three examples for the three types of flares. The top panel shows an example for a type I flare. One can see that both of the \lya\ and SXR emissions exhibit a good single peak and the \lya\ emission peaks more than one minute earlier than the SXR emission. Here we also plot the time derivative of the SXR flux that actually corresponds to the \lya\ emission very well. This indicates that the \lya\ emission in this flare follows a Neupert effect, which has already been reported before though for different flares. The middle panel displays an example for a type II flare. It is seen that the \lya\ emission reaches its maximum nearly at the same time as the SXR emission with a time difference of 8 s. In this case, the \lya\ emission obviously peaks later than the time derivative of the SXR emission. An example for a type III flare is shown in the bottom panel. We can see that the \lya\ emission rises slowly and reaches its maximum about three minutes later than the SXR emission. It should be highlighted that this case has rarely been reported before.

\subsubsection{Time Difference between the \lya\ and SXR Emission Peaks}

\begin{figure}[htb]
\epsscale{0.85}
\plotone{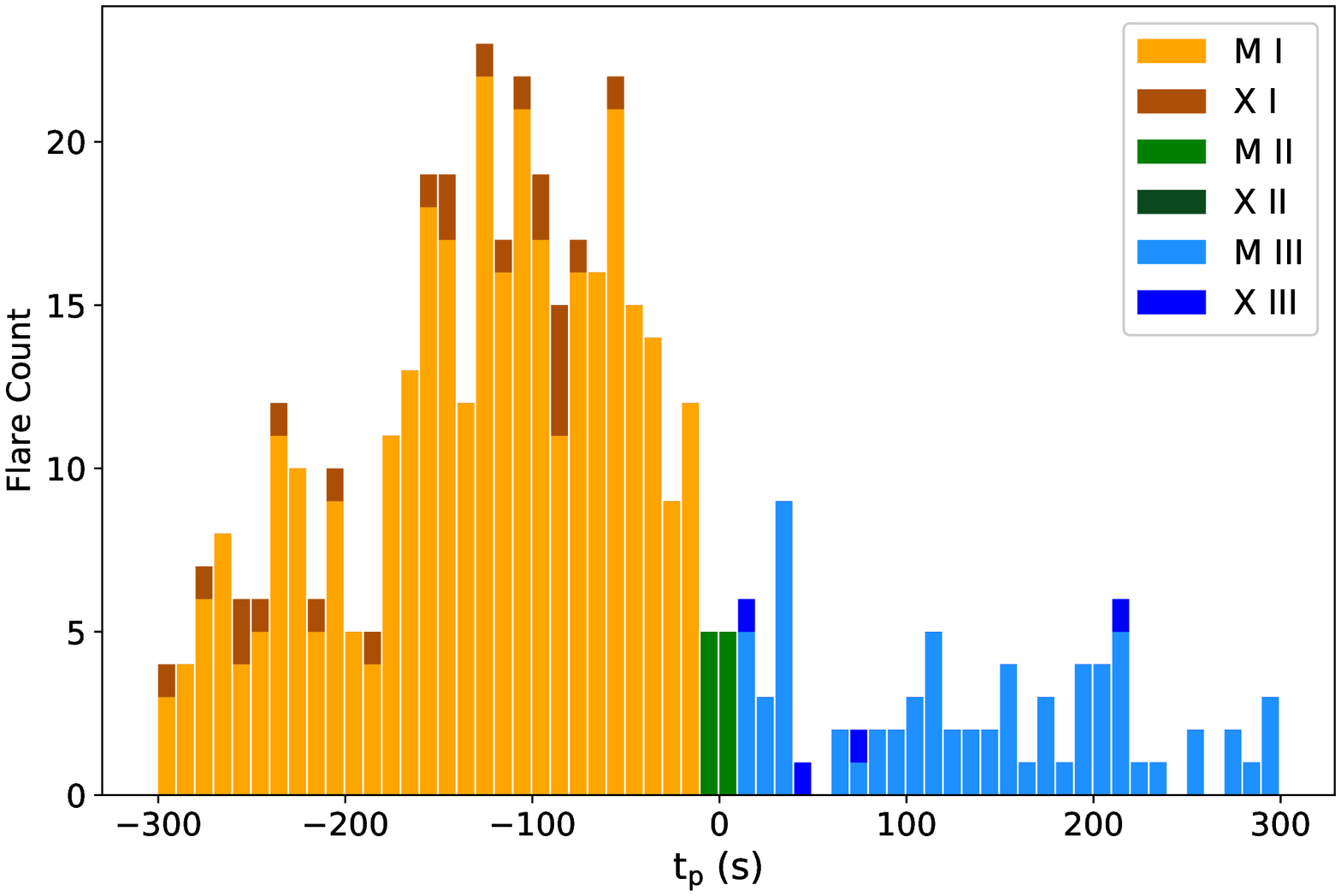}
\plotone{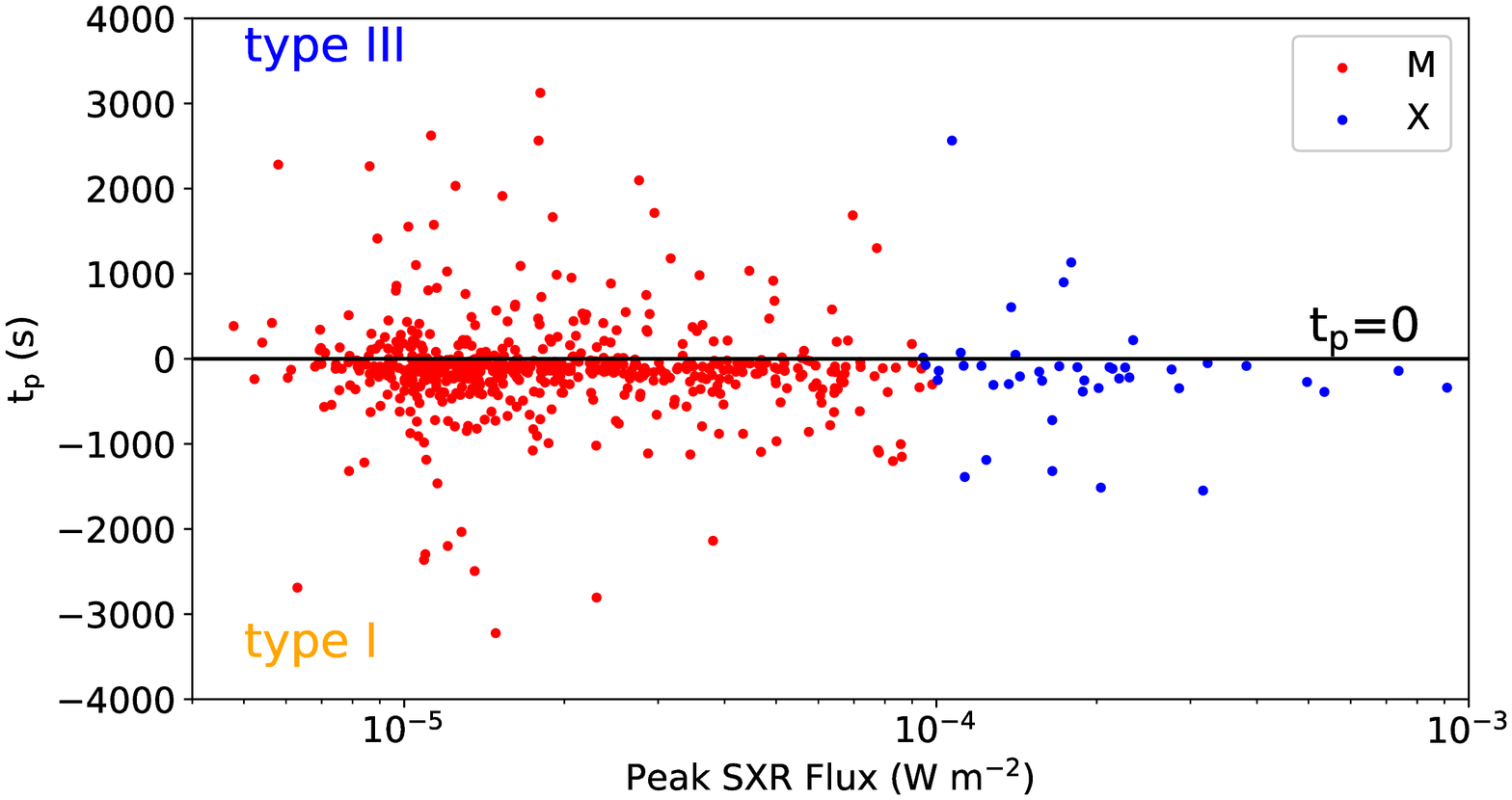}
\plotone{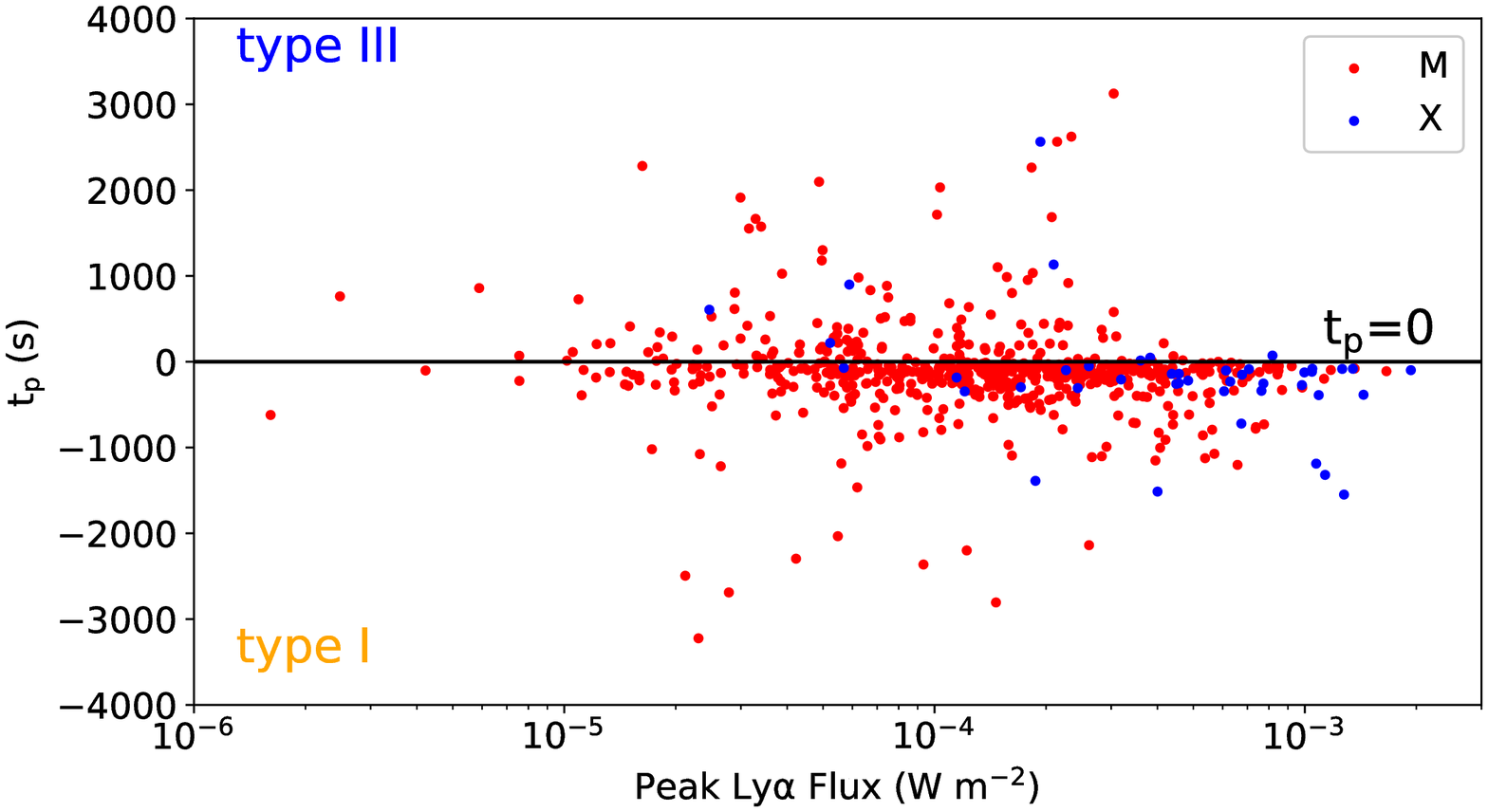}
\caption{Histogram of the time difference ($t_p$) between the \lya\ and SXR emission peaks (top) and scatter plots of $t_p$ versus peak \lya\ and SXR fluxes (middle and bottom). Note that the $t_p$ histogram has been cut to a range of $\pm$300 s for a better display. In the top panel, the orange, green, and blue colors denote the types I, II, and III flares, respectively, with the dark color marking X-class flares and the light color for M-class flares. In the middle and bottom panels, the red and blue dots represent the M- and X-class flares, respectively.}
\label{fig:tp}
\end{figure}

It is found that among the 658 flares in our dataset, there are 505 (76.8\%) type I flares that consist of 35 X-class and 470 M-class flares, 10 (1.5\%) type II flares all of which are M-class flares, and 143 (21.7\%) type III flares that consist of 8 X-class and 135 M-class flares. Here we compute the time difference (denoted by $t_p$) between the \lya\ and SXR emission peaks for all the three type flares and show the results in Figure \ref{fig:tp}. It should be mentioned that we only consider the \lya\ peak that falls into the range between the start and end times of each flare. This would actually underestimate the $t_p$ value for some type III flares that have a \lya\ peak later than the flare end time (see the two flares as shown in the bottom panels of Figure \ref{fig:typeIII-lc1to4}). After carefully checking the light curves, we find that this kind of type III flares are quite few and do not substantially affect the statistical results here. According to our definition, type I flares have negative values of $t_p$ and type III flares have positive values of $t_p$. Considering that the majority of flares (more than two thirds) have a value of $t_p$ within five minutes, here we only show the $t_p$ histogram in the range of $\pm$300 s in the top panel of Figure \ref{fig:tp}. It is seen that both types I and III flares include M- and X-class events, as already revealed from above. In addition, most of the type I flares have a $t_p$ within three minutes, while the $t_p$ of type III flares is distributed equally within five minutes in general. The middle and bottom panels of Figure \ref{fig:tp} show the scatter plots of $t_p$ versus peak SXR and \lya\ fluxes, respectively. It can be seen that there seems to be no obvious relationship between $t_p$ and the peak fluxes of \lya\ or SXR.

\subsubsection{Dependence on Flare Duration, Flare Location, and Solar Cycle}

\begin{figure}[htb]
\includegraphics[height=8cm]{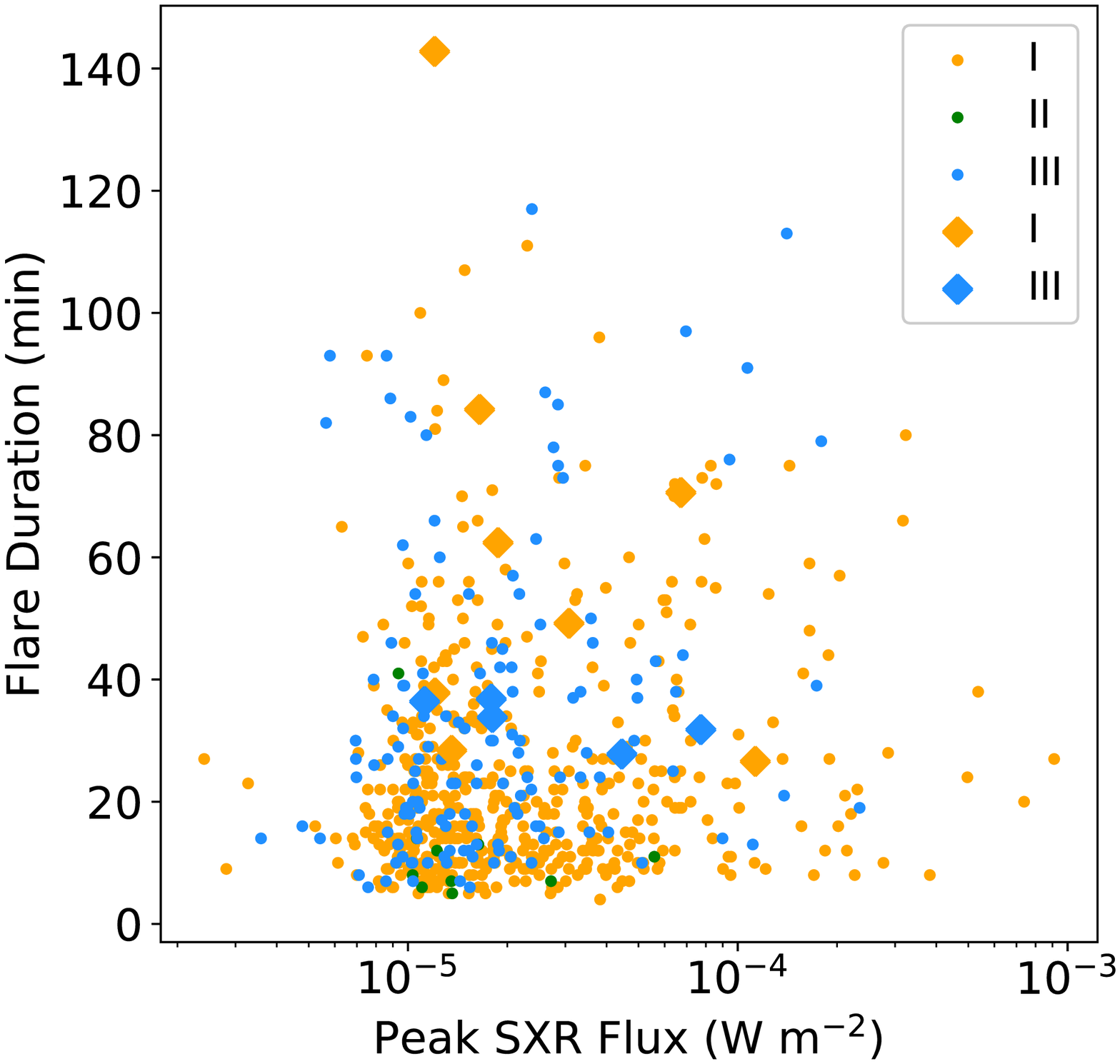} 
\includegraphics[height=8cm]{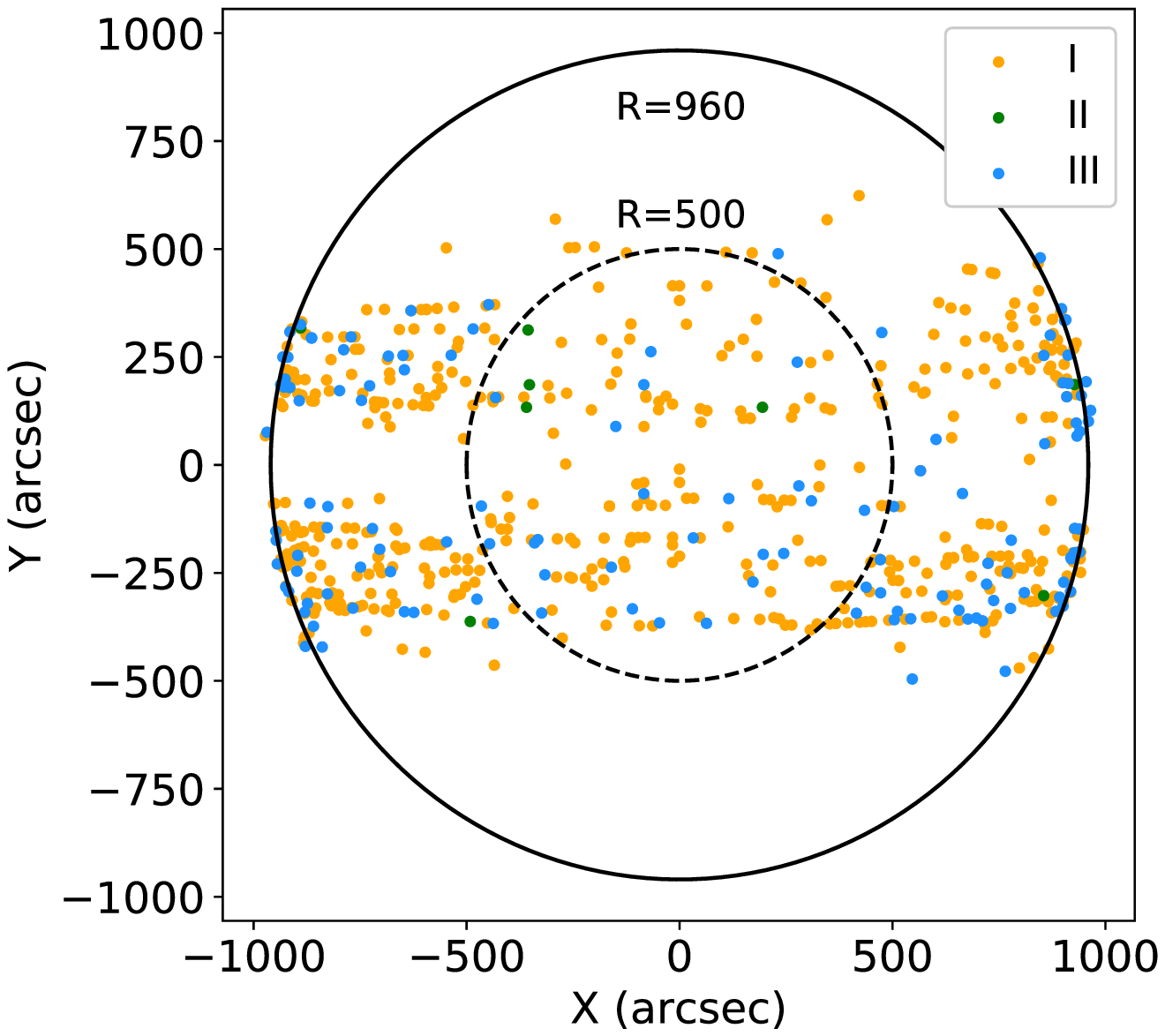}\\
\centering
\includegraphics[width=17.6cm]{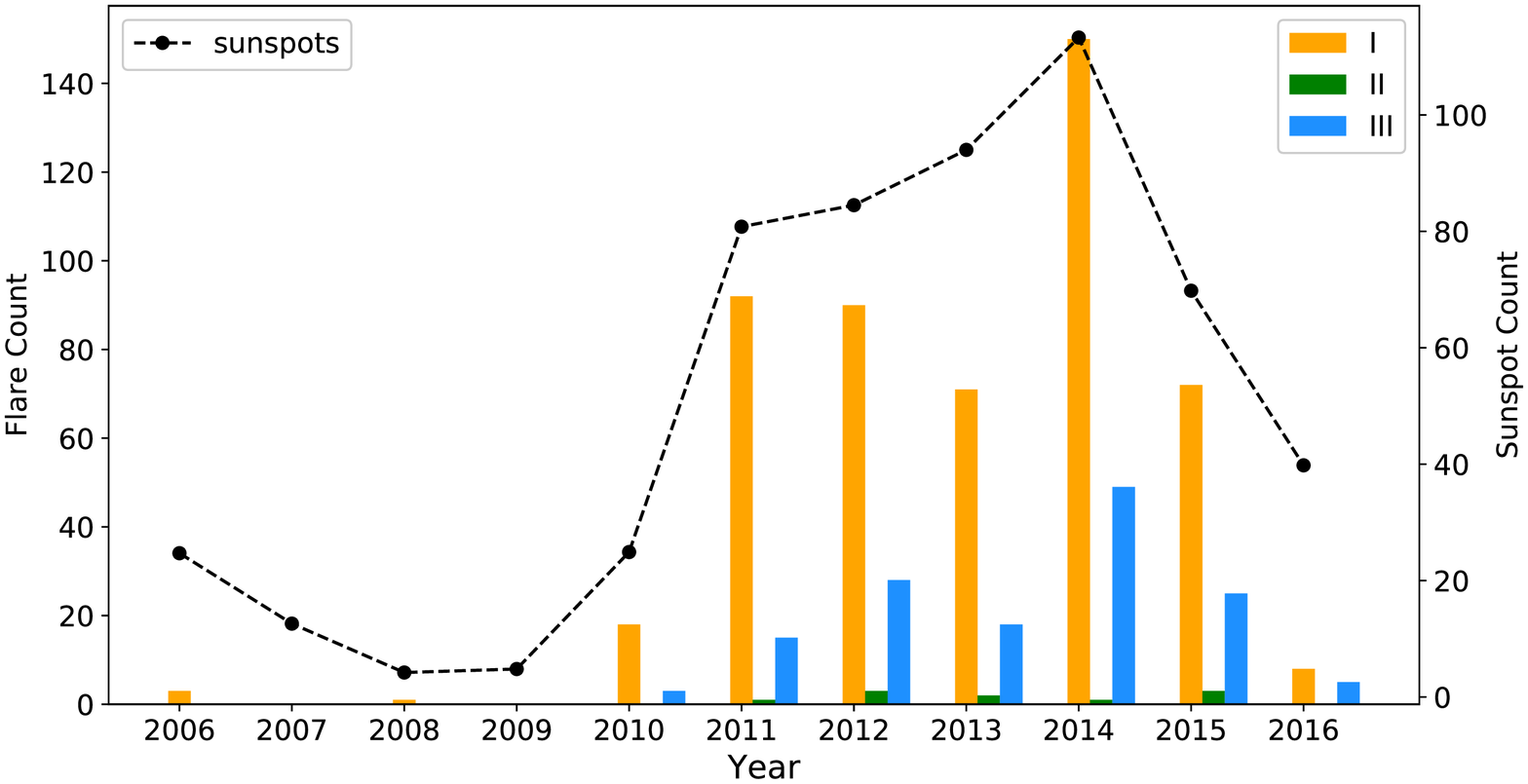}
\caption{The top left panel shows the scatter plot of flare duration versus peak SXR flux. Note that the diamonds represent flare durations that have been divided by five just for a better display. The top right panel plots the spatial distribution of flares on the solar disk with the solid circle ($R=960$) marking the solar limb. The dashed circle ($R=500$) marks the disk range in which we select 35 flares for loop length measurements. The bottom panel displays the variation of the flare counts and also the sunspot counts over years or solar cycle. In each of the panels, the orange, green, and blue colors denote the types I, II, and III flares, respectively.}
\label{fig:dist}
\end{figure}

We also check the dependence of flare types, more accurately speaking, of types I and III owing to quite a small number of type II flares, on the flare duration, location, and solar cycle, which is shown in Figure \ref{fig:dist}. From the top left panel we see that the flares with different types can be short- and long-duration flares, namely the flare type has no dependence on the flare duration. We can also see that the flare type is independent on the flare magnitude, either. From the spatial distribution of the flares with different types on the solar disk (the top right panel of Figure \ref{fig:dist}), one can see that there is no correlation between the flare type and the location. In addition, we plot the count variation of different type flares with years or solar cycle in the bottom panel of Figure \ref{fig:dist}, also overplotted the curve for sunspot counts. It is seen that the counts of flares with different types generally match the sunspot counts over the solar cycle.

\subsection{Multi-peak Feature of the \lya\ Emission in Types I and III Flares}

\subsubsection{Representative Light Curves}

\begin{figure}
\centering
\includegraphics[height=5cm]{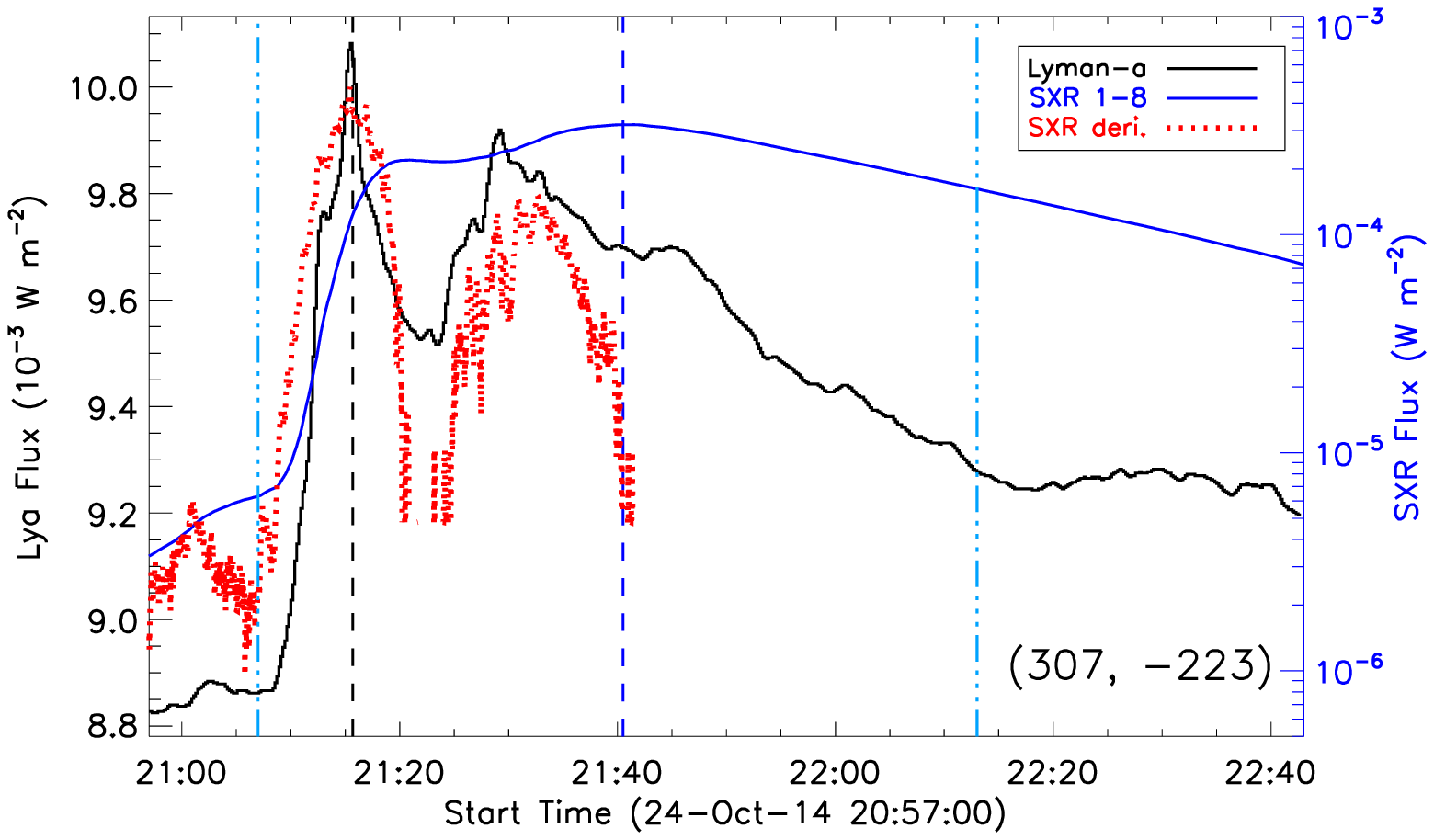}
\includegraphics[height=5cm]{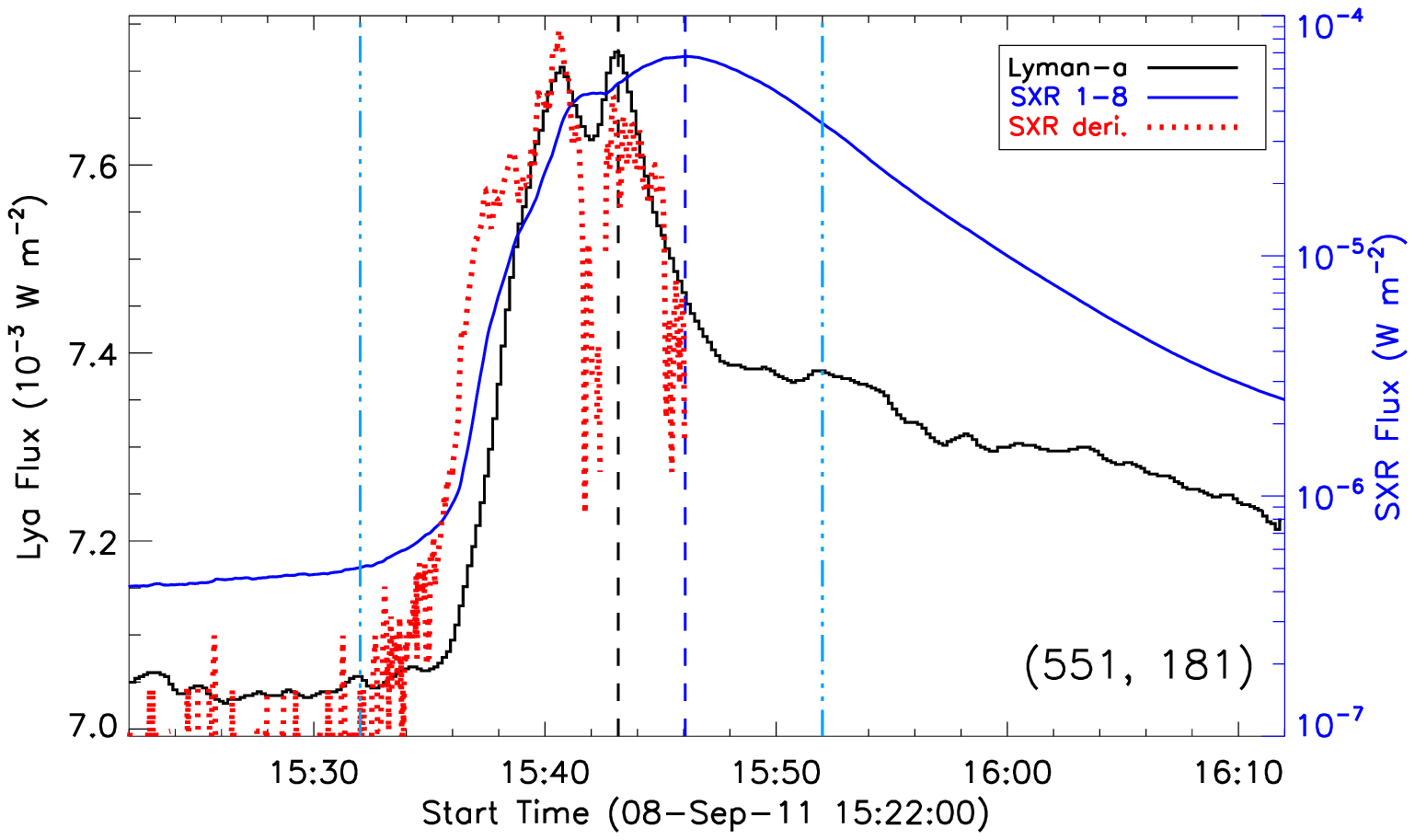}\\
\includegraphics[height=5cm]{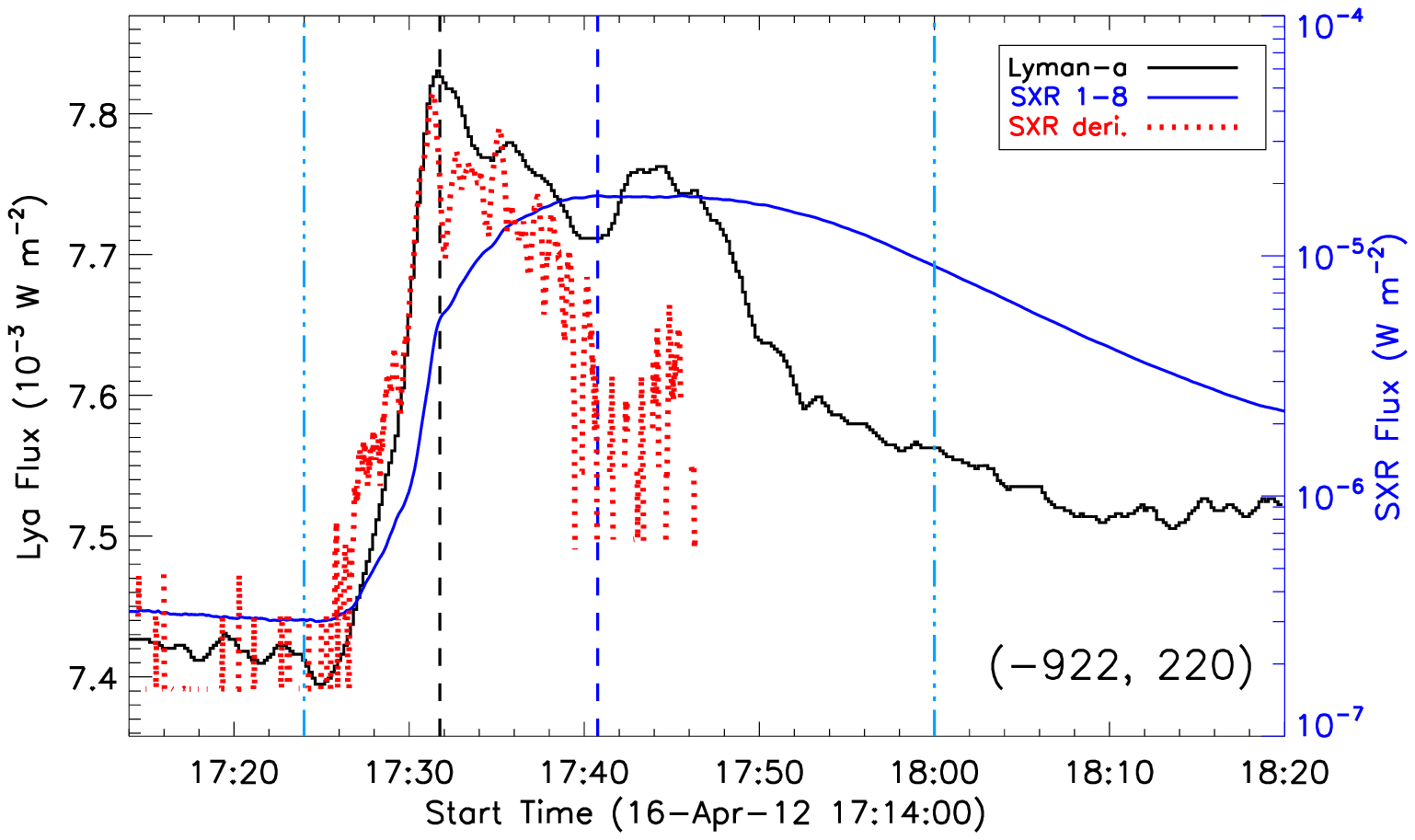}
\includegraphics[height=5cm]{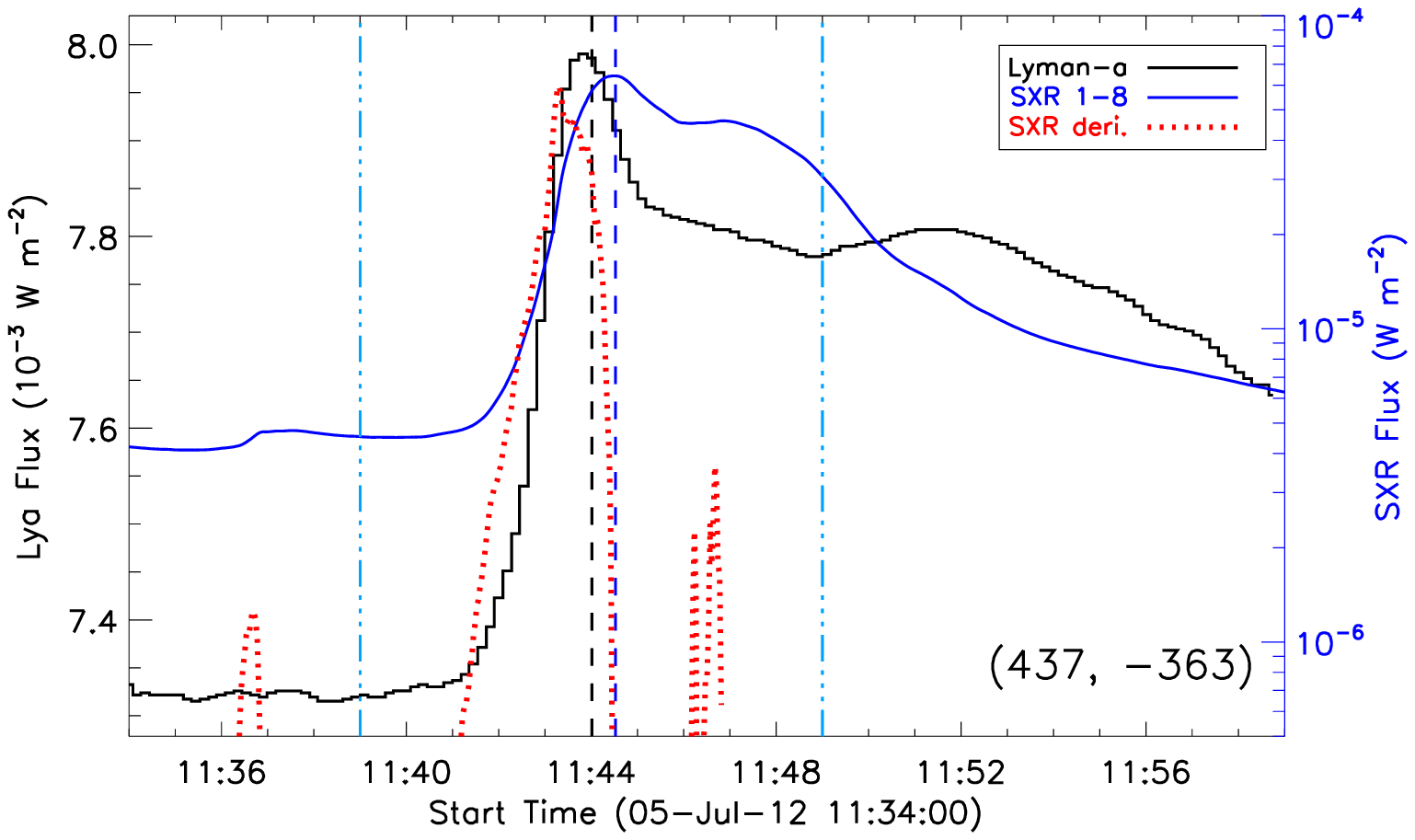}
\caption{Some representative light curves for type I flares showing a multi-peak feature in the \lya\ emission. All the curves and lines have the same meanings as the ones in Figure \ref{fig:types-lc}.}
\label{fig:typeI-lc1to4}
\end{figure}

\begin{figure}[htb]
\centering
\includegraphics[height=5cm]{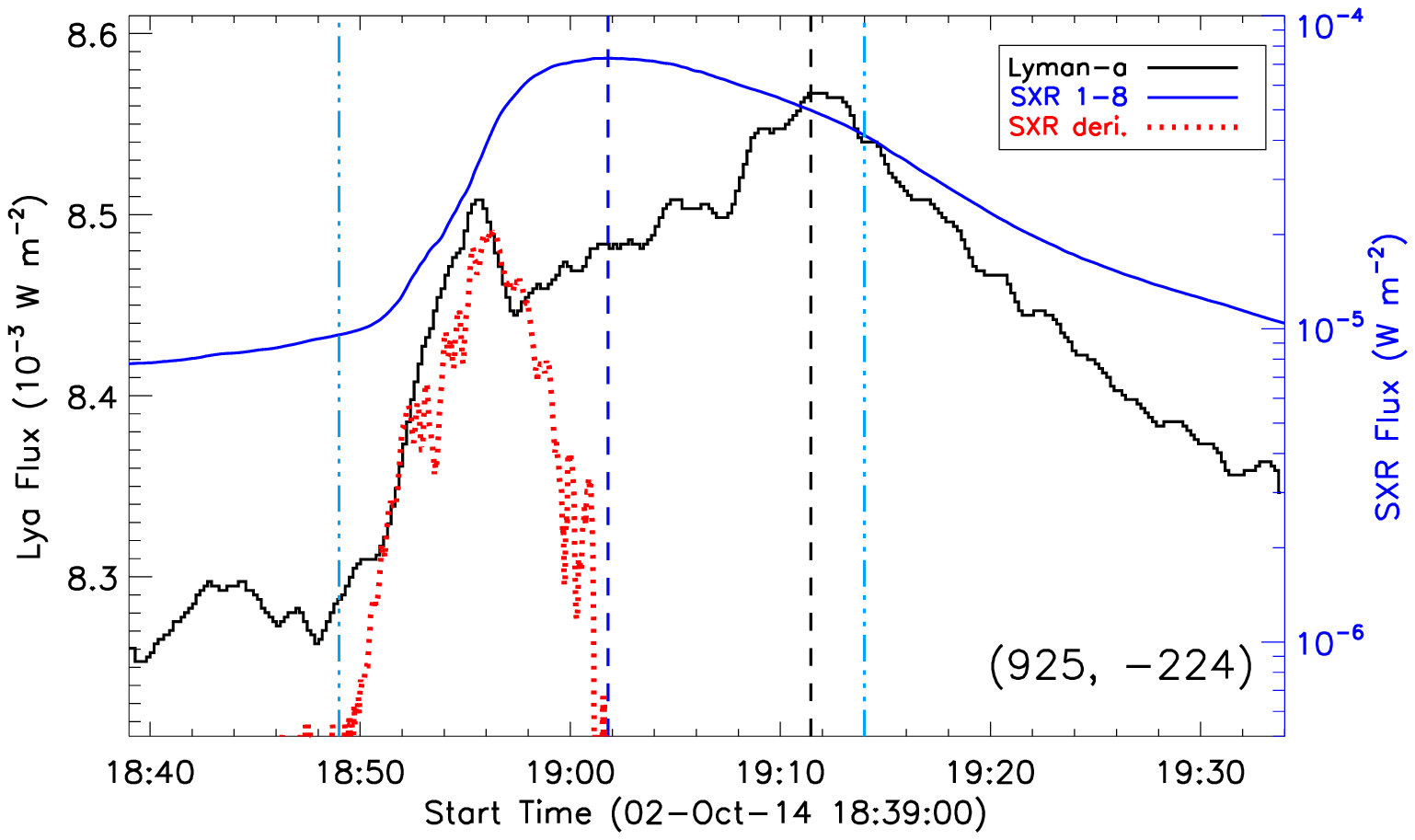}
\includegraphics[height=5cm]{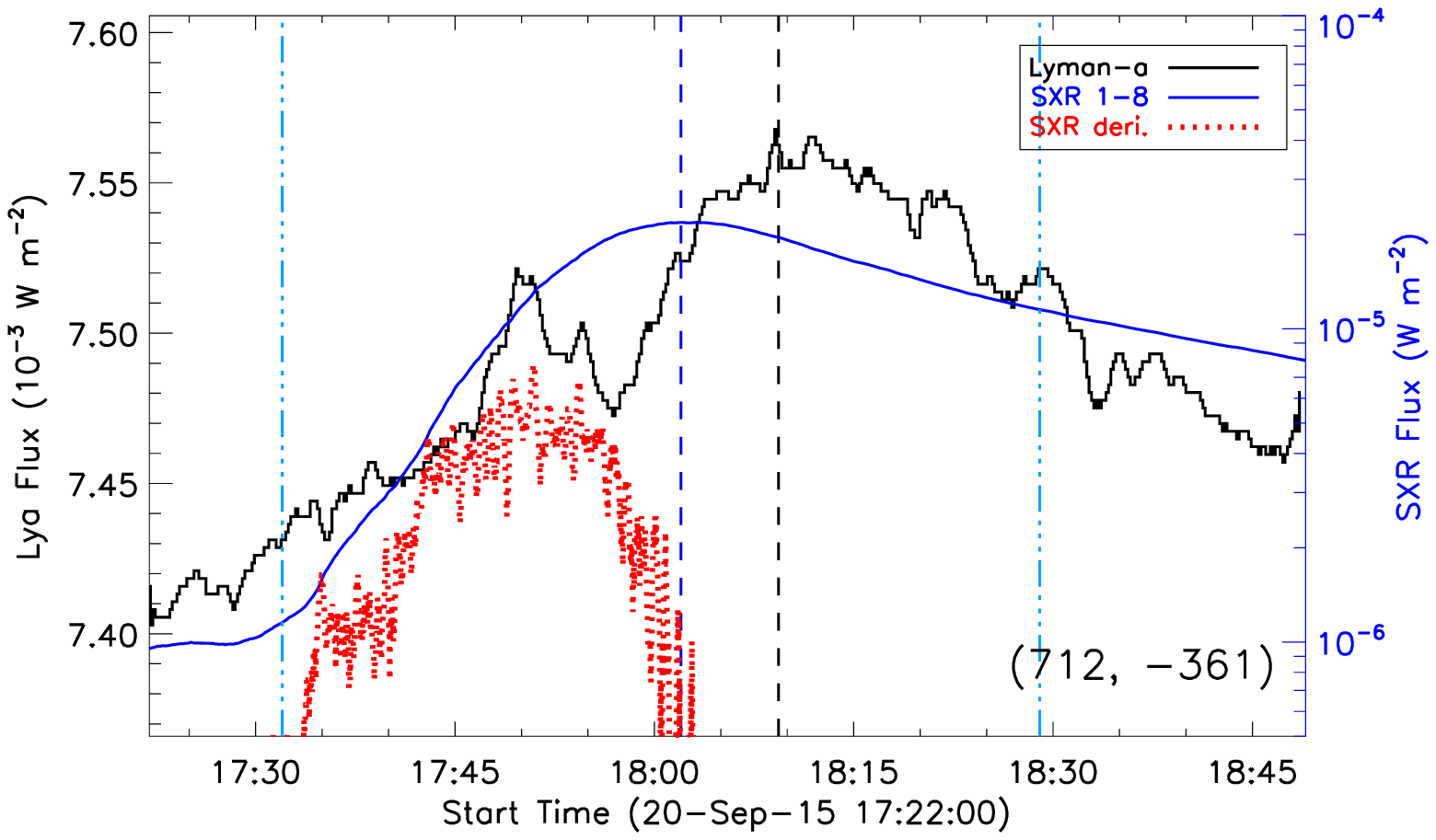}\\
\includegraphics[height=5cm]{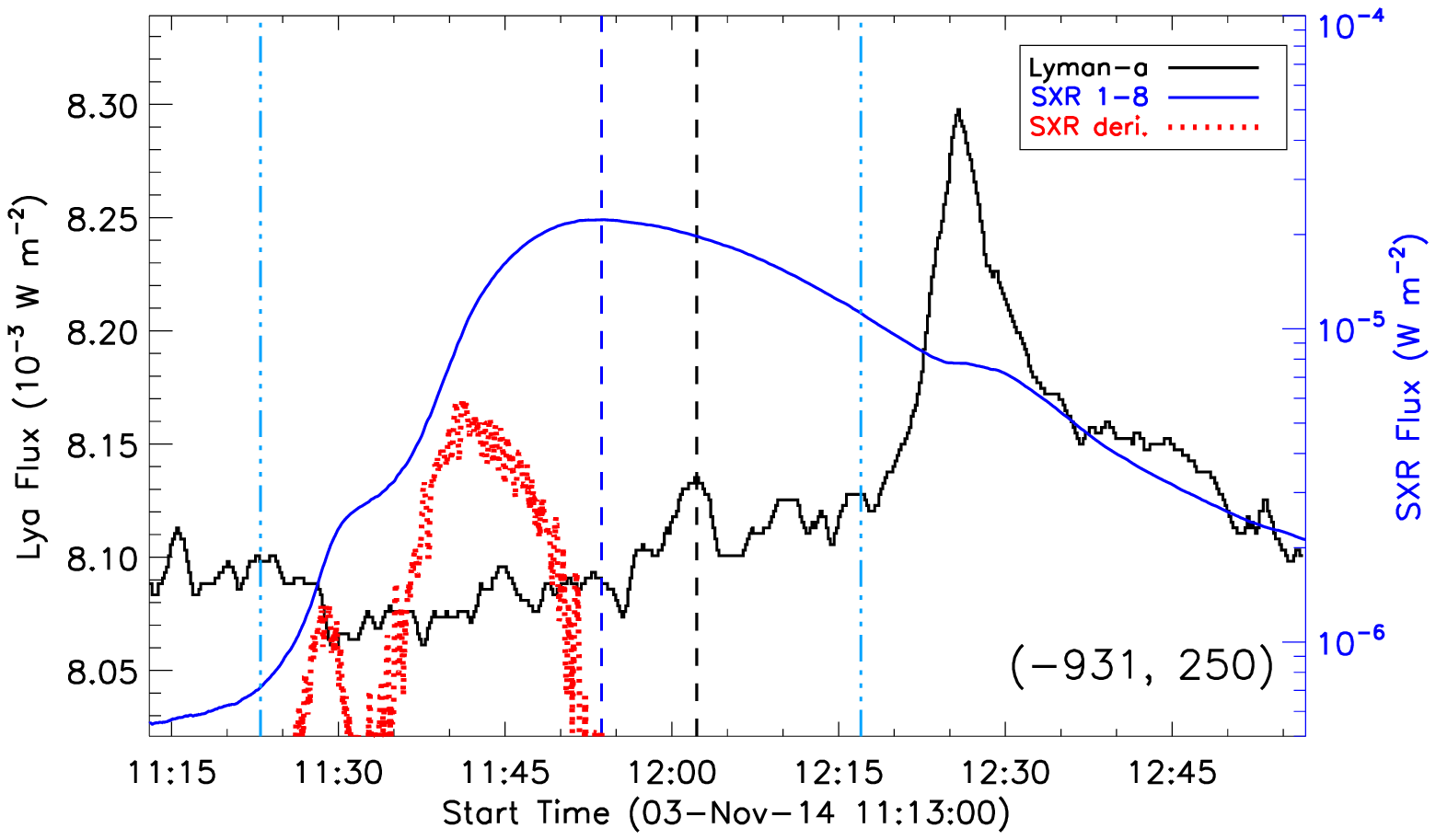}
\includegraphics[height=5cm]{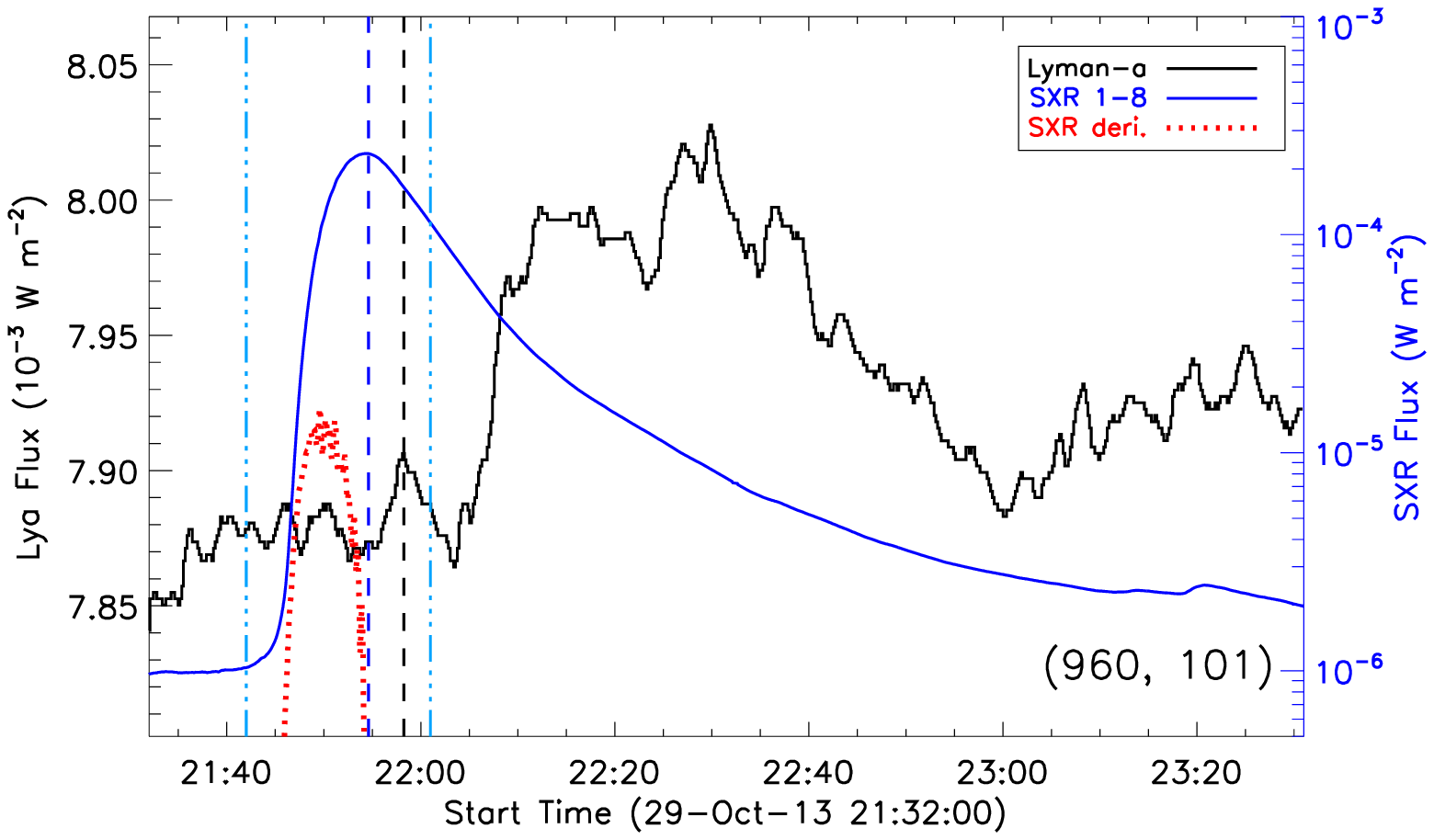}
\caption{Some representative light curves for type III flares showing a multi-peak feature in the \lya\ emission. All the curves and lines have the same meanings as the ones in Figure \ref{fig:types-lc}.}
\label{fig:typeIII-lc1to4}
\end{figure}

Apart from main peak, the \lya\ emission curve shows some evident sub-peaks in a large number of flares (roughly two thirds checked by eye), either for type I or type III. Such kind of light curves are plotted in Figures \ref{fig:typeI-lc1to4} and \ref{fig:typeIII-lc1to4}. From Figure \ref{fig:typeI-lc1to4} one can see that the \lya\ emission shows up one or two evident peaks, i.e., main peak and/or sub-peak, in the impulsive phase in the four type I flares, which resemble the time derivative of the SXR flux. Moreover, additional sub-peak appears in the gradual phase, with a time delay relative to the SXR peak. For the type III flares as shown in the top panels of Figure \ref{fig:typeIII-lc1to4}, the \lya\ emission exhibits a sub-peak in the impulsive phase which also matches the time derivative of the SXR flux, in addition to the main peak that appears in the gradual phase. Note that for the two type III flares in the bottom panels of Figure \ref{fig:typeIII-lc1to4}, the \lya\ emission does not show any significant enhancements in the impulsive phase but displays a few peaks in the gradual phase. In particular, a stronger peak appears in the late gradual phase. We notice that these two flares are located at or close to the solar limb (see the coordinates in the right corner of the panel), whose loop footpoints may be occulted (or partly occulted) by the solar disk. This would cause a weak \lya\ emission increase during the impulsive phase of the flare \citep{mill20}. While the significant enhancement of \lya\ in the late phase could be contributed by the flare loops that are still visible above the solar limb. Generally speaking, there usually exhibit an impulsive-phase peak (i.e., main peak for type I flares but sub-peak for type III flares) and a gradual-phase peak (main peak for type III flares but sub-peak for type I flares) in the \lya\ emission. Here it should be mentioned that there are also about one third flares which only show an isolated peak (or main peak) in \lya\ during the flare period. This case will be discussed in Section \ref{sec:summary}.

\subsubsection{Neupert Effect Check for the Impulsive-phase Peak of \lya}

\begin{figure}[htb]
\epsscale{0.85}
\plotone{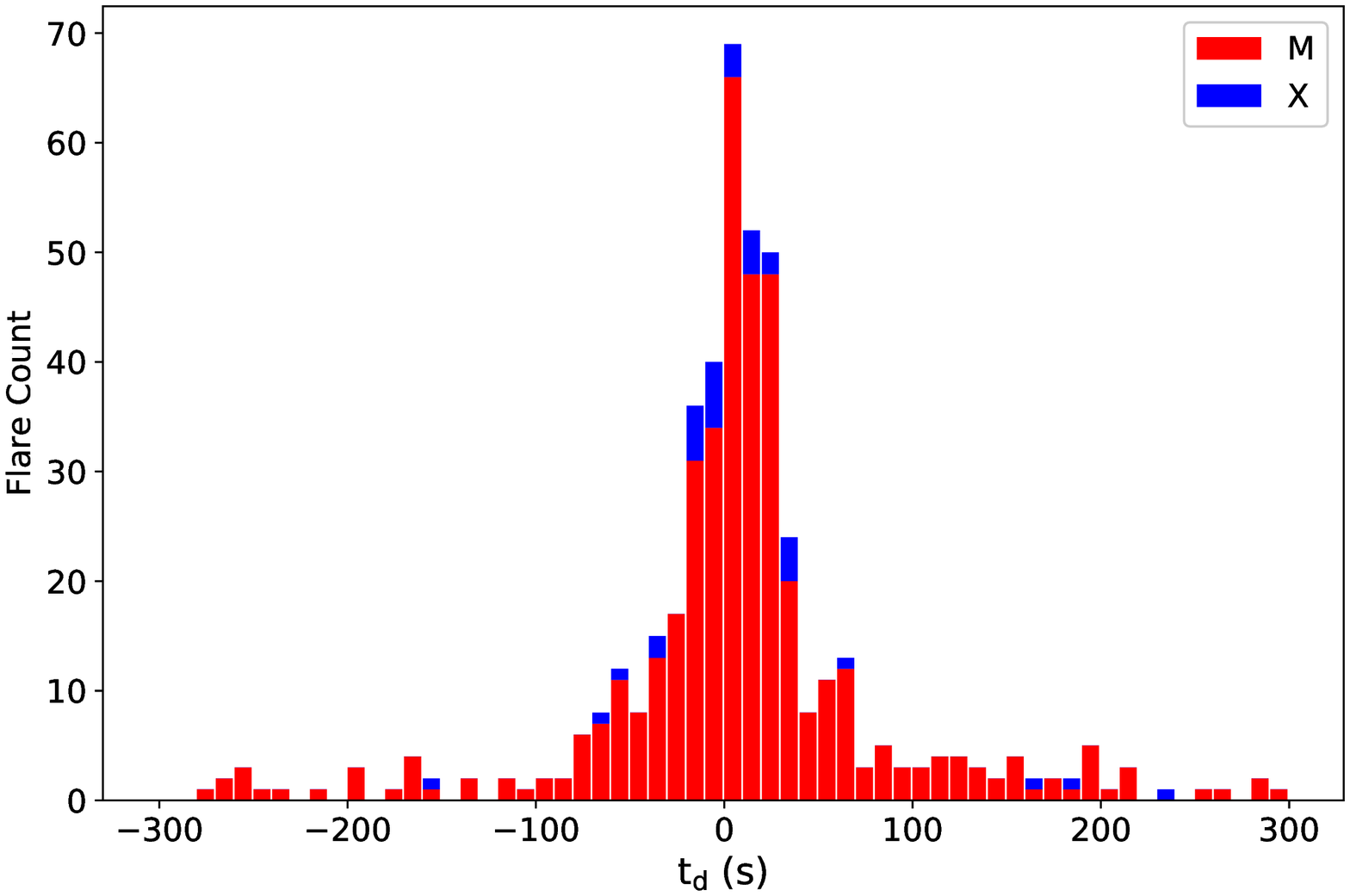}
\plotone{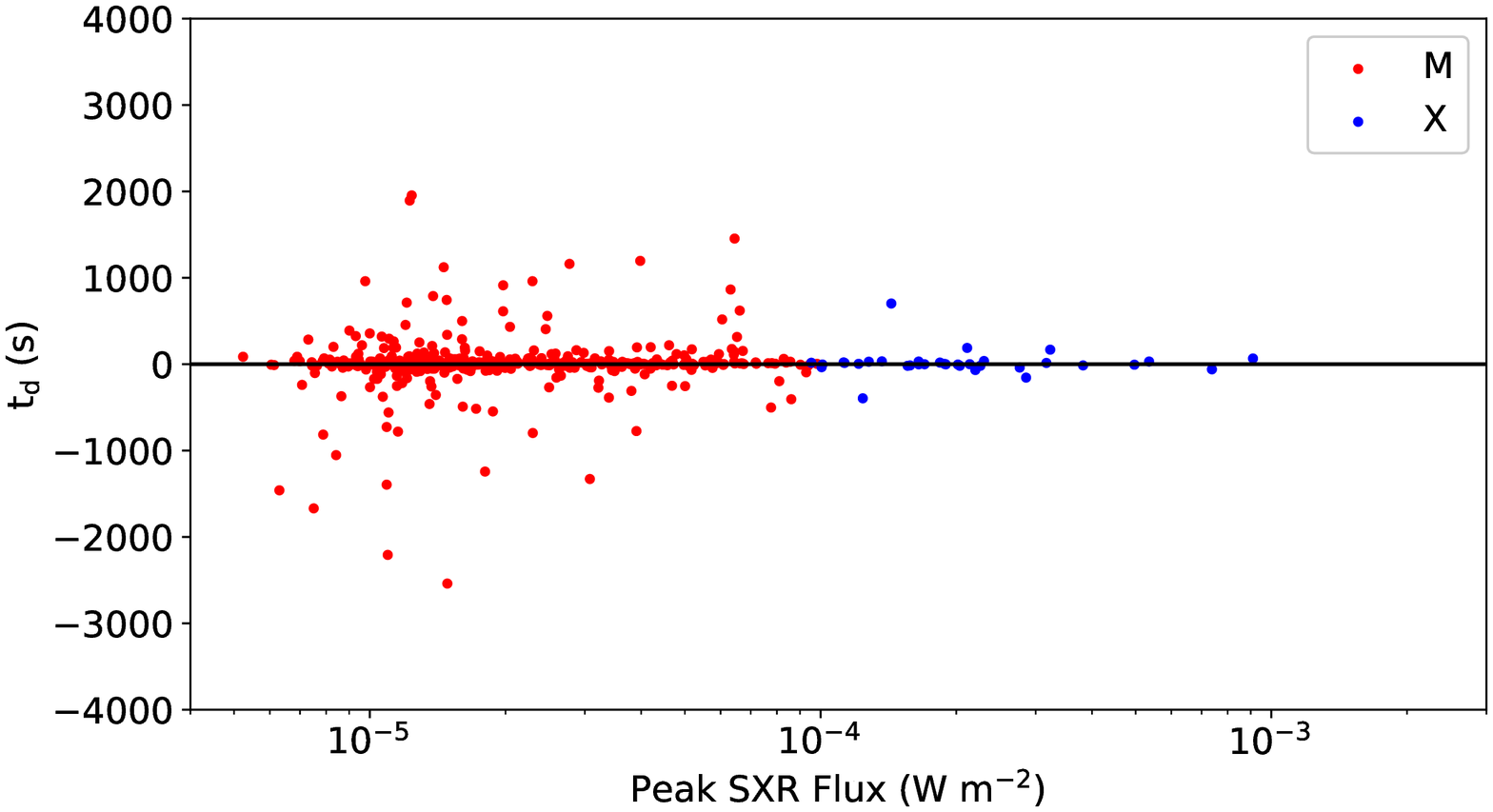}
\plotone{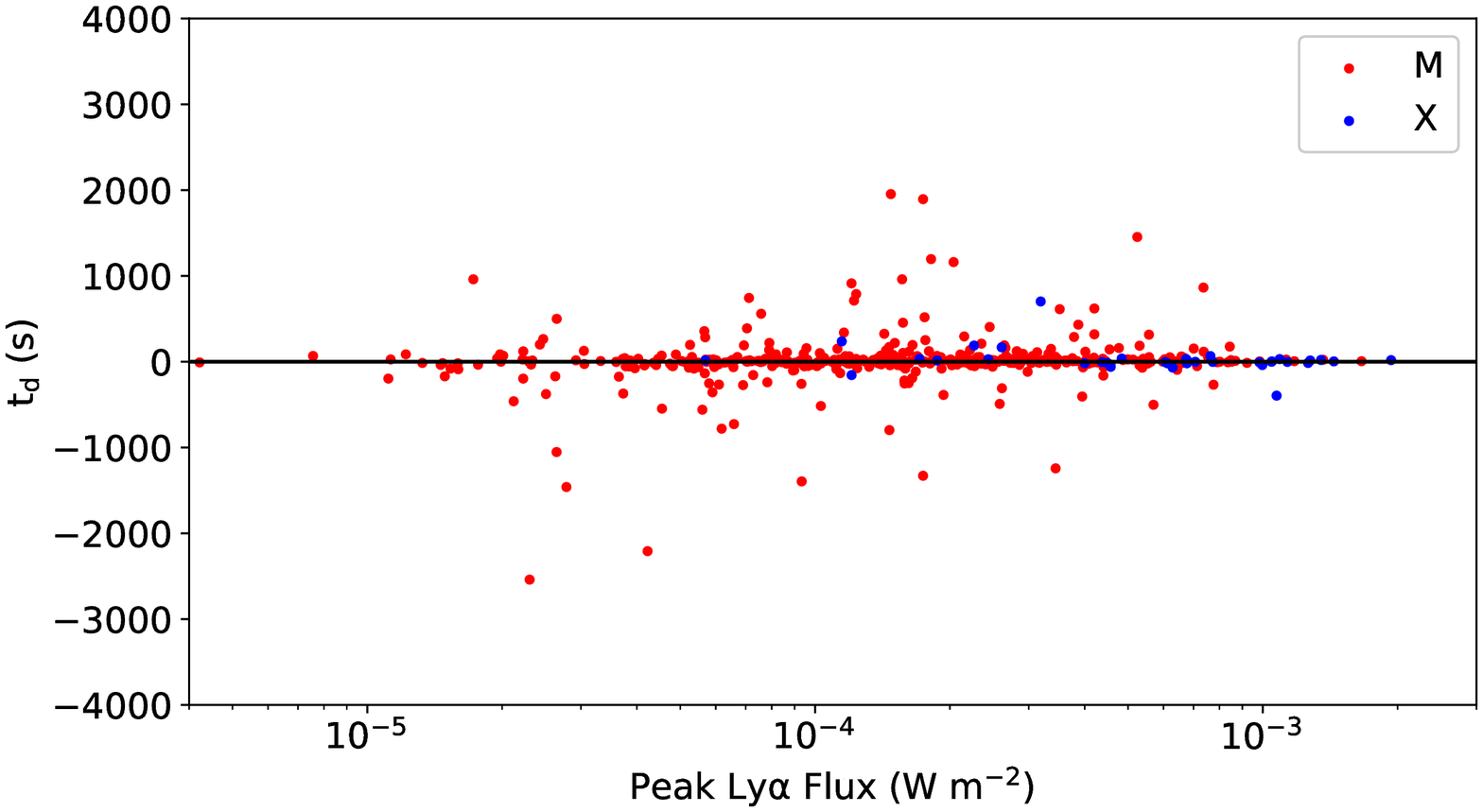}
\caption{Histogram of the time difference ($t_d$) between the \lya\ main peak and the peak of the time derivative of the SXR flux (top) and scatter plots of $t_d$ versus peak \lya\ and SXR fluxes (middle and bottom) for all of the type I flares. Note that the $t_d$ histogram has been cut to a range of $\pm$300 s for a better display. In each of the panels, the red and blue colors represent the M- and X-class flares, respectively.}
\label{fig:td}
\end{figure}

For the impulsive-phase peak emission of \lya, it is worthwhile to check its validity of the Neupert effect. First, we take a look at the main peak of \lya\ for all the 505 type I flares. The top panel of Figure \ref{fig:td} shows the histogram of the time difference (denoted as $t_d$) between the \lya\ main peak in the impulsive phase and the peak of the time derivative of the SXR flux. We can see that $t_d$, either for X- or M-class flares, has a notable peak around zero and the majority of flares ($\sim$80\%) have a $t_d$ value within $\pm$2 minutes. This indicates that the main peak emission of \lya\ in type I flares holds the Neupert effect generally. The middle and bottom panels of Figure \ref{fig:td} display the $t_d$ distribution versus the peak SXR and \lya\ fluxes, respectively. It is seen that larger flares have a smaller $t_d$ value, which may suggest that larger flares hold the Neupert effect better. However, there should be a caution that this might be owing to a data bias, since we have much fewer X-class flare examples than the M-class ones in our dataset. Regarding $t_d$ versus the peak \lya\ flux, it seems that there is no correlation between them.

\begin{figure}[htb]
\epsscale{0.8}
\plotone{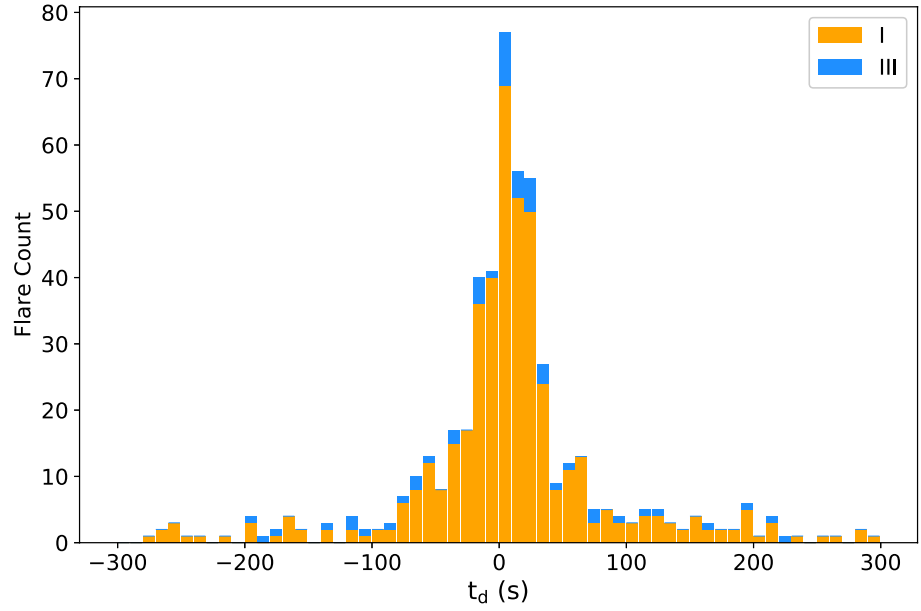}
\caption{Histogram of the time difference ($t_d$) between the impulsive-phase peak of \lya\ and the peak of the time derivative of the SXR flux for all of the type I flares (orange) and part (62) of the type III flares (blue).}
\label{fig:td-all}
\end{figure}

Furthermore, we identify an evident impulsive-phase peak (sub-peak) of \lya\ in 62 type III flares by eye and add their $t_d$ distribution on the result of type I flares (see Figure \ref{fig:td-all}). It is seen that similar to the main peak of type I flares, the impulsive-phase peak of \lya\ in type III flares also holds the Neupert effect mostly. Note that for the remaining 81 type III flares, it is somewhat hard to identify a well isolated impulsive-phase peak of \lya\ by eye owing to its very weak emission or too many small bumps in the \lya\ light curve.

\subsubsection{Delayed Time Check for the Gradual-phase Peak of \lya}

\begin{figure}[htb]
\epsscale{0.8}
\plotone{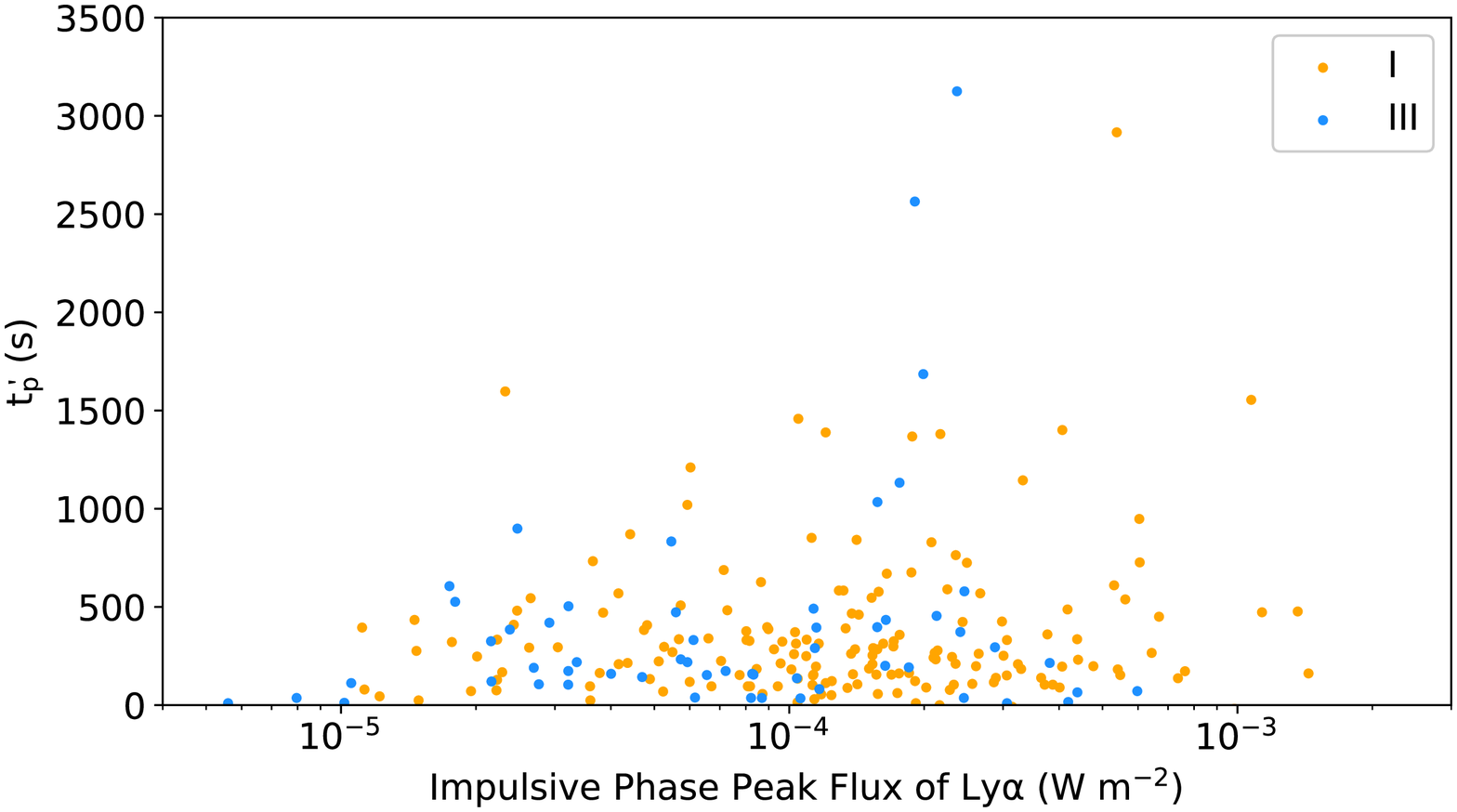}
\plotone{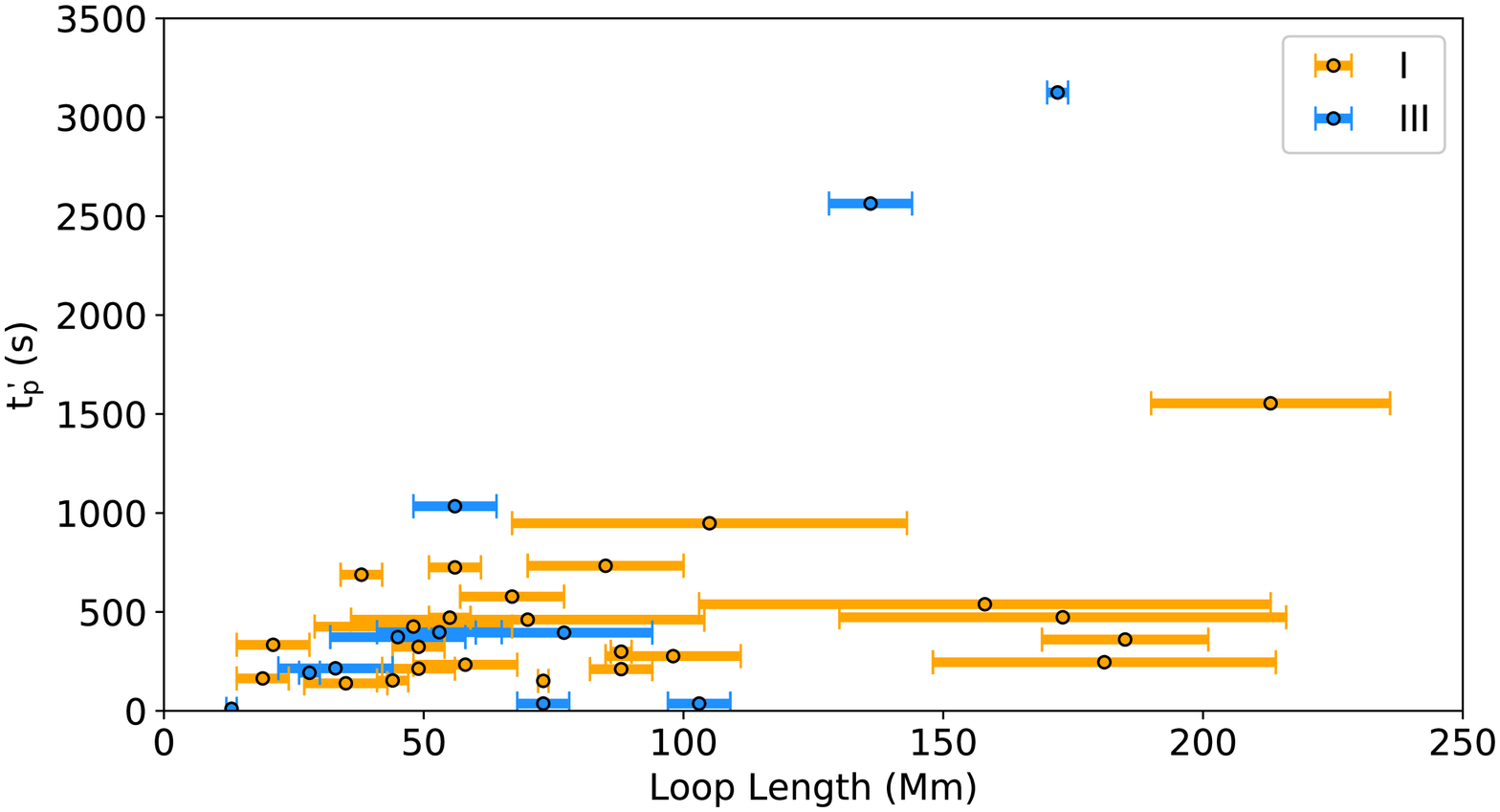}
\caption{The top panel shows the scatter plot of the delayed time ($t'_p$) of the gradual-phase peak of \lya\ versus the impulsive-phase peak flux of \lya\ for 173 type I flares (orange) and 62 type III flares (blue) in which both flare phase peaks can be well identified by eye. These two parameters have a very weak correlation with a coefficient of 0.1. The bottom panel shows the scatter plot of the delayed time ($t'_p$) of the gradual-phase peak of \lya\ versus the flare loop length for 24 (out of 173) type I flares (orange) and 11 (out of 62) type III flares (blue) that are located within a circle of a radius of 500\arcsec\ on the solar disk. There appears to be a correlation between the two parameters, with a coefficient of 0.5.}
\label{fig:tp-part}
\end{figure}

We also study the delayed time relative to the SXR peak for the gradual-phase peak of \lya\ and its relationship with the impulsive-phase peak flux of \lya\ and the flare loop length which are expected to be related to the initial heating and loop cooling \citep[e.g.,][]{carg95,yosh96}. Here we firstly identify a good gradual-phase peak (sub-peak) of \lya\ in 173 type I flares by eye and plot their delayed time (denoted as $t'_d$) versus the impulsive-phase peak (main peak) flux of \lya\ in the top panel of Figure \ref{fig:tp-part}. Furthermore, we over-plot the result for the 62 type III flares (delayed time of the main peak, or their $t_p$, versus the flux of the sub-peak of \lya\ in the impulsive phase) that are mentioned above. It is seen that there shows a cone shape or a weak correlation with a coefficient of 0.1 between the two parameters. Or we could say that as the impulsive-phase peak flux of \lya\ increases, the delayed time of the gradual-phase peak will more likely be larger.

In the bottom panel of Figure \ref{fig:tp-part}, we give the relationship of the delayed time for the gradual-phase peak of \lya\ with the flare loop length. Note that here we only select 35 flares (24 out of 173 type I flares plus 11 out of 62 type III flares) that are located within the circle of a radius of 500\arcsec\ (see the top right panel of Figure \ref{fig:dist}) to show this relationship. In this circle, the flare loop length could be measured more accurately by using AIA images due to the projection effect. Here we derive the loop length mainly from the AIA 171 \AA\ images combining with the AIA 1600 \AA\ images for the loop footpoint identification. Note that the horizontal error bar for the loop length comes from different sets of flare loops with different lengths in a single flare. From the selected 35 flares we can see that there appears to be a correlation, with a coefficient of 0.5, between the delayed time of the gradual-phase peak of \lya\ and the loop length.

\begin{figure}[htb]
\epsscale{0.85}
\plotone{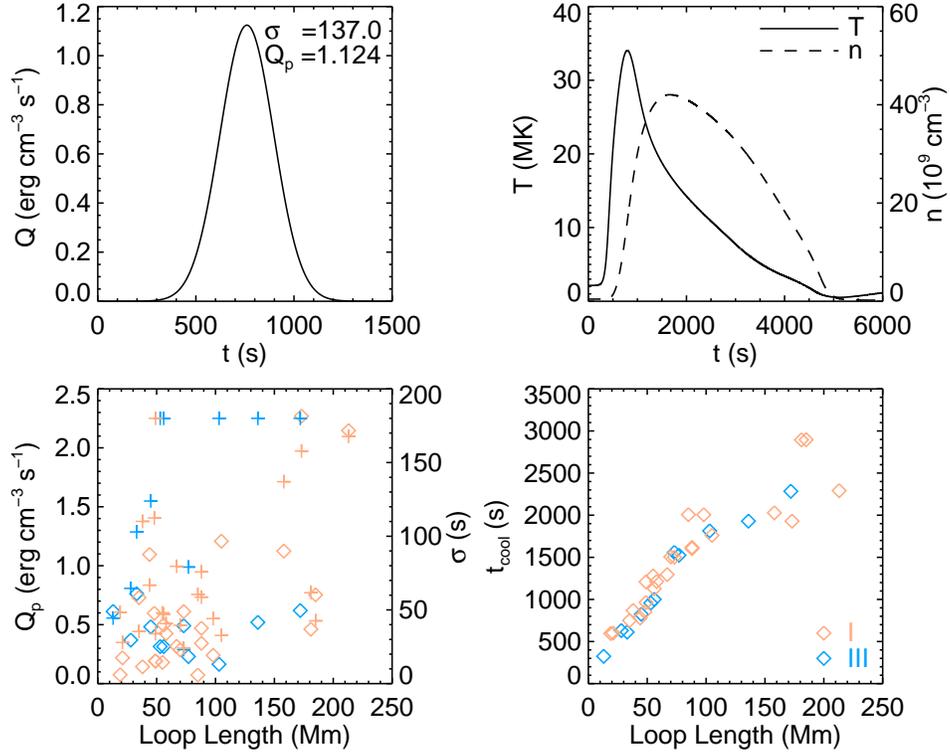}
\caption{EBTEL modelings for the 24 type I flares (orange) and 11 type III flares (blue) that are the same ones as shown in the bottom panel of Figure \ref{fig:tp-part}. The top left panel shows an example of the heating pulse ($Q$) with a Gaussian shape. The top right panel plots the time evolution of the average temperature ($T$) and election density ($n$) of a single loop for an example. The bottom left panel gives the peak heating rate ($Q_p$, diamond symbols) and the heating duration ($\sigma$, plus symbols) used in the modeling for the 35 types I (orange) and III (blue) flares. Here the flare loop length is measured from AIA images. The bottom right panel plots the relationship of the modeled cooling time ($t_{cool}$) with the loop length.}
\label{fig:ebtel}
\end{figure}

To further check the relationship between the delayed time of the gradual-phase peak of \lya\ and the impulsive-phase peak flux of \lya\ or the flare loop length, we carry out very simplified simulations for the selected 35 flares through the ``enthalpy-based thermal evolution of loops" (EBTEL) model \citep{klim08,carg12a,carg12b}. Here we treat each of the flares with a single-loop flare and the heating pulse assumed to be a Gaussian shape is constrained from the \lya\ light curve. More specifically, the peak heating rate (denoted as $Q_p$) and the heating duration (or the Gaussian width, $\sigma$) are proportional to the impulsive-phase peak flux of \lya\ and the flare rise time (i.e., from the flare onset time to the SXR peak time), respectively. We give an example for the heating pulse in the top left panel of Figure \ref{fig:ebtel}. The peak heating rate and the heating duration for all the 35 flares are shown in the bottom left panel, as a function of the loop length. It is seen that in our modeling, $Q_p$ (diamond symbols) is in the range of 0--2.5 erg cm$^{-3}$ s$^{-1}$ and $\sigma$ (plus symbols) is from 0--180 s. Note that here we set a maximum of $\sigma=$ 180 s in the single loop for a few flares that have a relatively longer rise time, according to some flare heating modelings via EBTEL \citep{raft09,ying12,ying14a}. The top right panel of Figure \ref{fig:ebtel} shows an example for the evolution of average temperature ($T$) and electron density ($n$) of a single loop. Using the average temperature, electron density, and loop length, we compute the cooling time ($t_{cool}$) and show its relationship with the loop length in the bottom right panel of Figure \ref{fig:ebtel}. Here $t_{cool}$ is defined as $1/t_{cool}=1/t_c+1/t_r$, where $t_c$ and $t_r$ are conductive and radiative cooling times, respectively \citep{carg95,raft09,carg12b,sunx13}. One can see that as the loop length increases, the cooling time becomes longer, which is consistent with the relationship between the delayed time of the gradual-phase peak of \lya\ and the flare loop length in observations as shown in the bottom panel of Figure \ref{fig:tp-part}. This suggests that the delayed time is a good indicator of plasma cooing, which is related to the flare heating magnitude and especially the flare loop length.


\section{Summary and Interpretation}
\label{sec:summary}

The statistical results as well as light curve features for \lya\ from our flare dataset are summarized in Tables \ref{tab:sum1} and \ref{tab:sum2}. According to the fact that the flare types have no dependence on the flare magnitude, duration, location, or peak fluxes of \lya\ or SXR, we conjecture that the \lya\ emission is likely contributed by some common processes in each of the flare, say flare heating and plasma cooling. In the following, we attempt to more specifically interpret the \lya\ emission mainly for type I and III flares based on its relationship with the SXR emission in the framework of the standard flare model \citep[e.g.,][]{raft09,holm11,huds11}.

\begin{table}[htb]
\begin{center}
\caption{Summary and Interpretation of the Results in the Study (Part I)}
\label{tab:sum1}
\begin{tabular}{cccc}
\tableline
\tableline
Flare types & ~~~~~~~~~~~~~~I & ~~~~~~~~~~~~II & III \\
         & ~~~~~~~~~~~~~~($t_p<-$10 s) & ~~~~~~~~~~~~($-$10 s $\leq$ $t_p\leq$ 10 s) & ($t_p>$ 10 s) \\
\tableline
Flare counts & ~~~~~~~~~~~~~~505 (76.8\%) & ~~~~~~~~~~~~10 (1.5\%) & 143 (21.7\%) \\
X-class counts & ~~~~~~~~~~~~~~35 & ~~~~~~~~~~~~0 & 8 \\
M-class counts & ~~~~~~~~~~~~~~470 & ~~~~~~~~~~~~10 & 135 \\
\tableline
Statistics & \multicolumn{3}{c}{$t_p$ has no dependence on peak fluxes of \lya\ or SXR.} \\
               & \multicolumn{3}{c}{The flare types are independent on flare magnitude, duration, or location.} \\
               & \multicolumn{3}{c}{The flare counts in each type match the sunspot counts over solar cycle.} \\
\tableline
{Interpretation} &  \multicolumn{3}{c}{The \lya\ emission is likely contributed by some common processes in flares.}\\
\tableline
\tableline
\end{tabular}
\end{center}
\end{table}

\begin{table}[htb]
\begin{center}
\caption{Summary and Interpretation of the Results in the Study (Part II)}
\label{tab:sum2}
\begin{tabular}{ccc}
\tableline
\tableline
 Multi-peak  & ~~~Impulsive-phase peak & ~~~Gradual-phase peak \\
of \lya & ~~~($t_d$) & ~~~($t'_p$) \\
 \tableline
Type I   & ~~~main peak & ~~~sub-peak \\
Type III & ~~~sub-peak  & ~~~main peak \\
\tableline
Statistics & ~~~$t_d<$ 2 min for $\sim$80\% flares & ~~~$t'_p$ VS impulsive-phase peak flux: cc=0.1 \\
               & ~~~Neupert effect valid generally  & ~~~$t'_p$ VS flare loop length: cc=0.5 \\
               &                              & ~~~confirmed by EBTEL modeling \\
\tableline
Interpretation & ~~~electron beam heating & ~~~thermal plasma cooling \\
                                & ~~~(nonthermal origin)           & ~~~(thermal origin) \\                                 
\tableline
\tableline
\end{tabular}
\end{center}
\end{table}

The impulsive-phase peak (i.e., main peak for type I flares and sub-peak for type III flares) of \lya\ basically holds the Neupert effect, which suggests that this kind of emission is closely related to the nonthermal electron beams \citep{nusi06,rubi09,mill16}. It is known that in the standard flare model, the electron beams deposit their energy mostly in the chromosphere via Coulomb collisions, where the \lya\ line is mainly formed. The local plasma is heated and then we can see evident emission in the \lya\ waveband together with some other bands. In particular, the \lya\ emission in this case is supposed to come from the chromospheric footpoints of flare loops, which has been supported by {\em TRACE} imaging observations \citep{rubi09}. This nonthermal origin of \lya\ is also confirmed in radiative hydrodynamic simulations. With an electron beam heating model, \cite{yang20} calculated the time evolution of the \lya\ line intensity and found that the \lya\ intensity peaks nearly at the same time as the nonthermal heating rate.

By contrast, the gradual-phase peak (i.e., main peak for type III flares and sub-peak for type I flares) of \lya\ that has a time delay relative to the SXR peak is supposed to be mainly caused by the thermal plasma cooling from flare loops. During a flare, the chromospheric plasma is heated and fills in the flare loops. Thus we see prominent hot emission in the SXR and also EUV wavebands. As the flare evolves, the hot plasma in flare loops suffers from a conductive or radiative cooling that usually takes place in the gradual phase. When the thermal plasma cools down to the formation temperature of the \lya\ line, we could see evident emission enhancement in the \lya\ waveband. This scenario is supported by our observational and modeling results, namely the delayed time of the gradual-phase peak of \lya\ ($t'_p$) or the plasma cooling time ($t_{cool}$) is related to the flare heating magnitude (represented by the impulsive-phase peak flux of \lya) as well as the flare loop length. In fact, cool flare loops with a chromospheric temperature have been observed in recent years \citep[e.g.,][]{hein18,koza19,hein20}. In addition, \cite{mill20} reported that the \lya\ energy is comparable to, or about an order of magnitude smaller than the total thermal energy. This implies that the energy from thermal plasma is enough to provide the emission that is radiated by \lya.

Note that in some of the flares, the \lya\ emission can show multiple sub-peaks in the impulsive phase or in the gradual phase in addition to a main peak. The former case suggests that there probably exist multiple heatings during the flare, which likely take place in multiple loop strands \citep[e.g.,][]{hock12,ying14a,reep18,reep19,reep20}. While the latter case indicates that such kind of flares are probably EUV late-phase flares characteristic of different sets of loops with different lengths (e.g., \citealt{wood11,hock12,sunx13,ying14b,wood14}). Considering that a real flare consists of multiple loop strands (also implied from the error of the loop length measured from AIA images in the present study), multiple heatings/coolings are supposed to exist in a series of loop strands and cause a relatively long rise/decay or multiple peaks in the \lya\ light curve.

It should also be noted that there are about one third flares whose \lya\ emission does not show a well isolated impulsive-phase peak (i.e., sub-peak for type III flares) or a well identified gradual-phase peak (i.e., sub-peak for type I flares) but only an evident main peak (like the one shown in the top panel of Figure \ref{fig:types-lc}). One possibility is that these \lya\ sub-peaks are too weak to be identified by eye or to be submerged in the background emission. Another possibility is that these \lya\ sub-peaks overlap with the main peak and can not be well separated from the main peak, probably due to a long time heating (say, proceeding into the decay phase) or a very short time cooling (say, with a very short loop length). 

Overall, the \lya\ emission can be originated from both of a nonthermal electron beam heating and a thermal plasma cooling. These two processes are supposed to play a different role in contributing the \lya\ emission in different types of flares. We conjecture that the nonthermal electron beam heating may play a major role in contributing the \lya\ emission in type I flares that have a relatively larger impulsive-phase peak of \lya. While the thermal plasma cooling may be more important in type III flares whose gradual-phase peak of \lya\ is greater.


\section{Conclusion and Discussions}
\label{sec:dis}

In this paper, we have performed a statistical study on the relationship of the \lya\ emission with the 1--8 \AA\ SXR emission in solar flares. It is worth highlighting that there are about one fifth (type III) flares whose \lya\ emission peaks later than the SXR emission. This result is complementary to the prior studies on flaring \lya\ that usually showed an earlier maximum in \lya\ than in SXR (i.e., type I flares). Based on the Neupert effect check as well as the delayed time analysis and modeling, we conclude that the \lya\ emission in different types of flares could be of nonthermal origin as well as thermal origin. 

The nonthermal origin of \lya\ has been illustrated in previous studies \citep[e.g.,][]{nusi06,rubi09,mill16}, though only with case studies for a few large flares. Our statistical results in the present work provide a further confirmation for that. Note that here we just use the time derivative of the SXR flux rather than HXR emissions to serve as the evidence for nonthermal electron beams due to unavailable HXR observations for numerous flare examples. Regarding the thermal origin of \lya, we owe it to a delay of the \lya\ peak with a few to tens of minutes relative to the SXR peak, which is further confirmed from the relationship between the delayed (or cooling) time and the flare loop length. In fact, it is natural and reasonable that the \lya\ emission can come from the coronal loop cooling \citep[e.g.,][]{mill20}. This can be predicted in the well-established standard flare model based upon decades of flare analyses and modelings. Here we would like to draw attention that the thermal origin of \lya\ can be of importance in some cases (say, in type III flares) and that the role of nonthemal and thermal processes may change from flare to flare, which could yield different types of \lya\ emission curves. Note that we cannot rule out a thermal (or direct) heating that could also be contributed to the \lya\ emission, as the radiative hydrodynamic simulations showed by \cite{hong19}. Considering that a statistical study cannot touch upon all of the features of the \lya\ emission, we propose that some more detailed case studies are needed to further explore the physical property of the flaring \lya\ particularly with comprehensive observations from multiple instruments. For instance, the HXR observations from {\em RHESSI} can provide more clear evidence for the flare heating, i.e., nonthermal electron beams or not \citep[e.g.,][]{holm11}. The spectral observations at multiple temperatures from {\em SDO}/EVE can be used to better trace the cooling pattern of thermal plasma \citep[e.g.,][]{thie17}. Moreover, the imaging observations from {\em SDO}/AIA can give us the spatial information on both the flare loops and loop footpoints.

The present study can improve our understanding on the physical nature of the \lya\ emission in solar flares, so as to help interpret the near-future \lya\ observations made by the {\em Solar Orbiter} \citep{mull19} and the {\em Advanced Space-based Solar Observatory} ({\em ASO-S}, to be launched around 2022; \citealt{ganw19}). {\em Solar Orbiter} carries an Extreme Ultraviolet Imager (EUI; \citealt{roch19}) that can observe the Sun closely at 0.28 AU in the \lya\ waveband with a high spatial resolution. The \lya\ Solar Telescope (LST; \citealt{lihu19}) on board {\em ASO-S} will obtain the full-disk images in \lya\ from a Sun-synchronous orbit. These spatially-resolved observations contain much more detailed information on the \lya\ emission features and the present study gives a preliminary basis for diagnosing the nonthermal and thermal origins of \lya\ in solar flares. In addition, the present study can provide some constraints for the radiative hydrodynamic simulation of \lya\ in solar flares, which is actually a powerful tool to interpret the optically-thick \lya\ line emission.


\acknowledgments
The authors thank the anonymous referee very much for the constructive comments/suggestions that significantly improve the manuscript. {\em SDO} is a mission of NASA’s Living With a Star Program. The authors are supported by NSFC under grants 11873095, 11733003, 11903020, 11961131002, and U1731241, and by the CAS Strategic Pioneer Program on Space Science under grants XDA15052200, XDA15320301, and XDA15320103. Y.L. is also supported by the CAS Pioneer Talents Program for Young Scientists.

\bibliographystyle{apj}


\end{document}